\documentclass[epj,nopacs,final]{svjour}
\usepackage{graphics}
\usepackage{latexsym,overpic,rotating,float}
\usepackage{subfigure,color,showkeys}
\usepackage{epsfig,color,rotating,amsmath,delarray,array,multirow}
\usepackage{makeidx,pifont,float,amssymb}
\usepackage{ulem} %% for strike-through
\newcommand{\er}{$\pm$}
\newcommand{\be}{\begin{eqnarray}}
\newcommand{\ee}{\end{eqnarray}}
\newcommand{\bea}{\begin{eqnarray}}
\newcommand{\eea}{\end{eqnarray}}
\newcommand{\bc}{\begin{center}}
\newcommand{\ec}{\end{center}}

\newcommand{\beq}{\begin{equation}}
\newcommand{\eeq}{\end{equation}}
\newcommand{\ba}{\begin{eqnarray}}
\newcommand{\ea}{\nonumber \end{eqnarray}}

\newcommand{\bi}{\begin{enumerate}}
\newcommand{\ei}{\end{enumerate}}
\def\MSL (#1,#2,#3,#4){[\Lambda\frac{#1}{2}^#2]_{#3}(#4)}
\def\MSS (#1,#2,#3,#4){[\Sigma\frac{#1}{2}^#2]_{#3}(#4)}

\begin{document}

\title{\boldmath Highlights of the Spectroscopy of Hyperons and Cascade Baryons}
\titlerunning{Highlights of the Spectroscopy of Hyperons and Cascade Baryons }

\author{Annika Thiel\inst{1} and Eberhard~Klempt\inst{1}\\}
\authorrunning{Annika Thiel and Eberhard~Klempt}

\institute{
\inst{1}Helmholtz--Institut f\"ur Strahlen-- und Kernphysik, Universit\"at Bonn, 53115 Bonn, Germany\\
}

\date{\today}

\abstract{We highlight topical issues in hyperon spectroscopy that will be accessible when a secondary
 beam of neutral kaons is created in Hall D at Jefferson Lab as proposed to the JLab Program Advisory
Committee. The new beam will have a flux of $10^4$ $K_L$/s. The reaction products will be analyzed in the
GlueX experimental setup. We point out which physical questions need to be answered and suggest a number
of novel experiments. In particular we suggest to measure the presently disputed SU(3) structure of the
$\Lambda(1405)$, to search for Pentaquarks mit exotic quantum numbers and to search for baryons 
belonging to the SU(6) 20-plet representation. }

%\pacs{25.75.-q}

\maketitle

\section{Introduction}
Our present knowledge on the spectrum of $\Lambda$ and $\Sigma$ hyperons and of cascade
baryons $\Xi$ still relies on experimental data taken in 1960s and 1970s. The masses, widths and
decay properties of light baryons carrying strangeness S $=-1$ or $-2$ derived from early analyses are
collected in the bi-annual Review of Particle Physics~\cite{Tanabashi:2018oca}. Recent re-analyses
of the old data have cleaned up the spectrum slightly and reported evidence for some new $\Lambda$ and $\Sigma$
hyperons~\cite{Zhang:2013cua,Zhang:2013sva,Fernandez-Ramirez:2015tfa,Kamano:2014zba,Kamano:2015hxa,Matveev:2019igl,Sarantsev:2019xxm}.

The search for new hyperon resonances and to confirm less-known ones is important to
establish the multiplet structure of excited baryons. In the sector of $N$ and $\Delta$ resonances
the first excitation band is completely known and well established, most states in the second
excitation band are at least seen -- even though some of them with fair evidence only --, ten
states (out of 45) in the third band are known, and $N$ and a $\Delta$ Regge trajectories
exist up to states with orbital angular momentum $L=6$. Our knowledge on the hyperon spectrum of $\Lambda$ and
$\Sigma$ resonances is much poorer:
even the first excitation band is not complete, and some states in the first band are known
with little evidence. Only few states in the second excitation band have been reported so far. No
Regge trajectory can be drawn with more than two states. 
Our knowledge on the cascade
resonances is even worse: Apart from the ground states (in the SU(3) octet and decuplet), only
one resonance has been reported with spin and parity.

In particular the $\Lambda$ states are sensitive to details of quark models. Consider the two spin doublets
with $J^P = 3/2^+$\,/\,$5/2^+$: $\Lambda(1890)$\,/\,$\Lambda(1820)$  and
$\Lambda(2070)$\,/\,$\Lambda(2110)$. The low-mass doublet is interpreted in all quark models as
SU(3) partner of $N(1720)$ /$N(1680)$. The high-mass doublet, however, is interpreted as partner of $N(1880)$/\linebreak$N(1900)$
in the celebrated Isgur-Karl model~\cite{Isgur:1978wd} that uses an effective gluon exchange for the quark-quark interaction,
in contrast to
the Bonn model \cite{Loring:2001ky} that is based on instanton-induced interactions. This model predicts the
$\Lambda(2070)$\,/\,$\Lambda(2110)$ doublet as SU(3) singlet states.
The latter two states have been reported in a recent Bonn-Gatchina analysis~\cite{Sarantsev:2019xxm}, with fair evidence only; the
decay modes seem to favor their SU(3) singlet nature. The example demonstrates the sensitivity of
the hyperon spectrum to the interaction between quarks in the confinement region.

Not only quark models are on the test bench. Modern approaches to hadron spectroscopy generate
resonances from their decay products. A famous example is the $\Lambda(1405)$ region which is supposed
to house two $\Lambda$ and one or two $\Sigma$ resonances with spin-parity $J^P=1/2^-$. The
well-known $\Lambda(1405)$ is seen with a large SU(3)-octet contribution, a new wider 
$\Lambda(1380)$ as largely SU(3) singlet. 
Quark models predict only one state in this mass region, the $\Lambda(1405)$ as SU(3) singlet state.
Alternative approaches will be discussed below.  In any case, sensitive experiments with high 
precison are required to resolve this conflict. 

After a period of great enthusism, the interest in searches for pentaquarks had degraded considerably
even though the discussion continued (see, e.g.~\cite{Kuznetsov:2017ayk}). The discovery 
of three $J/\psi p$ structures observed in 246.000 $\Lambda_{b} ^{0}\to J/\psi p K^-$ events 
by the LHCb collaboration, the pentaquark candidates $P_c(4312)$, $P_c(4440)$, and $P_c(4457)$~\cite{Aaij:2019vzc}, 
has reinitiated the
interest. A search the pentaquark candidates in the photoproduction reaction
$\gamma p\to J/\psi p$ with 469 $J/\psi$ events did not find evidence for a pentaquark
enhancement \cite{Ali:2019lzf}. 
In the light-quark sector, three pentaquarks are predicted with quantum numbers
that are not accessible to three-quark systems. These states are, of course, of particular interest
and should be searched for with high statistics and, in the case of narrow states, with good resolution.

These topics ask for a new facility allowing for the study of Kaon-nucleon interactions in the
mass range from threshold to 3\,GeV. Such a facility is offered by the proposal PAC47 at JLab~\cite{Amaryan:2017ldw}.
A beam of $K_L$-mesons will be provided with an intensity of $10^4$ particles per second.
This facility will provide for new insights into the spectroscopy of hyperons and cascade baryons.
Also the COMPASS experiment~\cite{Ketzer:2019wmd}, J-PARC~\cite{Sako:2013prop}, and the 
forthcoming PANDA experiment~\cite{Iazzi:2016fzb}
may provide substantial contributions to hyperon spectroscopy.

The paper is organized as follows. After this Introduction, the physics of hyperons in the first and third
excitation band -- the negative-parity states -- will be discussed in Section~ \ref{2nd},
including a proposal how to determine the SU(3) structure of the $\Lambda(1405)$ resonance. In Section~\ref{3rd}, the positive-parity
states in the second excitation band will be discussed as well as the possibility to search for
two of the three light-quark pentaquarks that have flavor quantum-numbers that are incompatible
with three-quark states. Three-quark baryons have two oscillators that can be excited. In most excitations,
only one of the two oscillators is excited (with rapid changes from one oscillator to the other one), or
there is at least a component in the baryon wave function in which only one oscillator is excited.
In quark models there is, however, also a class of resonances that always carry excitation in both oscillators.
Yet, no member of this class has ever been observed. A scenario in which such a state
should be observable is presented at the end of Section~\ref{3rd}. A short discussion of the physics of
Regge trajectories and of cascade baryons follows in Section~\ref{4th} and \ref{5th}.

\section{\label{2nd}The negative-parity states}
\subsection{The first and third excitation band}
Baryon resonances are often classified by the leading representation in the flavor-spin
SU(6) basis by $J^P (D,L^P_N) S$ where $J^P$ is the spin and parity of a resonance,
$D$ the dimensionalilty of the SU(6) representation, $L$ the intrinsic orbital angular momentum,
$N$ the excitation band, and $S$ the total quark spin.

The first excitation band has an intrinsic orbital angular momentum $L=1$; the resonances
in the first band belong to a $(D,L^P_N) = (70, 1^-_1)$ representation. The 70-plet can be
expanded into a spin-doublet SU(3) decuplet, a spin-doublet and a spin-quartet SU(3)
octet, and a spin-doublet SU(3) singlet:
\be
70\ =\ ^2\hspace{-1mm}\ 10\ \oplus\ ^2\hspace{-1mm}\ 8\ \oplus\ ^4\hspace{-1mm}\ 8\ \oplus\ ^21\,.
\ee

The $N$ and $\Delta$ resonances belonging to the first excitation band are all well
established. Resonances with a leading configuration with
spin and isospin 1/2 have a mass around 1530\,MeV; states
with a leading spin-1/2 and isospin-3/2 or spin-3/2 and isospin-1/2 configuration have masses that fall into a
(1660\er40)\,MeV window (see Table~\ref{star:N+D}).

The $\Lambda$ excitations in the first band are mostly also well established except for the state $J^P=3/2^-$ 
state that has dominantly internal quark spin $S=3/2$, see Table~\ref{star:lambda}.
This state was shown to have a very small $K^-N$ decay width~\cite{Guzey:2005rx,Guzey:2005vz}, it is hence difficult to find it
in $K^-p$ scattering. It may be more advantageous to search for the state in the decay sequence
$\Sigma^+({\rm high\ mass})\to \Lambda(xxx)3/2^-\pi^+\to(\Sigma\pi)\pi^+$.
Unfortunately, $N(1700)3/2^-$ was not yet seen as intermediate state in a cascade process,
hence no predictions can be made concerning the best suited cascade.

\begin{table}
\caption{\label{star:N+D}Star rating of nucleon and $\Delta$ resonances in the first excitation band.
$J^P$ is the spin and parity of a resonance, $S$ the total internal quark spin. The table gives
the leading configuration; mixing between states with identical external quantum numbers can mix.}
\bc
\renewcommand{\arraystretch}{1.35}
\begin{tabular}{||c|ccc|cc||}
\hline\hline
&\multicolumn{3}{c|}{Octet}&\multicolumn{2}{c||}{Decuplet}\\
    \hspace{4mm} $J^P=$   & $1/2^-$ & $3/2^-$& $5/2^-$& $1/2^-$& $3/2^-$\\\hline
$70\ [^210]$&&&&****&****\\[-3ex]
                       &&&&\tiny $N(1620)$ & \tiny $N(1700)$\\[-0.5ex]
$70\ [^48]$&****&****&****&&\\[-3ex]
                       &\tiny $N(1650)$ & \tiny $N(1700)$&\tiny $N(1675)$ &&\\[-0.5ex]
$70\ [^28]$&****&****&&&\\[-3ex]
                       &\tiny $N(1535)$ & \tiny $N(1520)$& &&\\
\hline\hline
\end{tabular}
\renewcommand{\arraystretch}{1.0}
\ec
\end{table}
\begin{table}
\caption{\label{star:lambda}Star rating of $\Lambda$ resonances in the first excitation band. See
caption of Table~\ref{star:N+D}.}
\bc
\renewcommand{\arraystretch}{1.35}
\begin{tabular}{||c|cc|ccc||}
\hline\hline
&\multicolumn{2}{c|}{Singlet}&\multicolumn{3}{c||}{Octet}\\
 \hspace{4mm} $J^P=$  & $1/2^-$ & $3/2^-$& $5/2^-$& $1/2^-$& $3/2^-$\\\hline
$70\ [^48]$&&&****&&****\\[-3ex]
                       &&&\tiny $\Lambda(1800)$ & \tiny - &\tiny $\Lambda(1830)$ \\[-0.5ex]
$70\ [^28]$&&&****&****&\\[-3ex]
                       &&&\tiny $\Lambda(1670)$ & \tiny $\Lambda(1690)$&\\[-0.5ex]
$70\ [^21]$&****&****&&&\\[-3ex]
                       &\tiny $\Lambda(1405)$ &\tiny $\Lambda(1520)$ &&&\\[-0.5ex]
\hline\hline
\end{tabular}
\renewcommand{\arraystretch}{1.0}
\ec
\end{table}
\begin{table}
\caption{\label{star:sigma}Star rating of $\Sigma$ resonances in the first excitation band. See
caption of Table~\ref{star:N+D}.}
\bc
\renewcommand{\arraystretch}{1.35}
\begin{tabular}{||c|ccc|cc||}
\hline\hline
&\multicolumn{3}{c|}{Octet}&\multicolumn{2}{c||}{Decuplet}\\
\hspace{4mm} $J^P=$   & $1/2^-$ & $3/2^-$& $5/2^-$& $1/2^-$& $3/2^-$\\\hline
$70\ [^210]$&&&&*&*\\[-3ex]
                       &&&&\tiny $\Sigma(1950)$ &\tiny $\Sigma(1920)$ \\[-0.5ex]
$70\ [^48]$&***&\tiny - &****&&\\[-3ex]
                       &\tiny $\Sigma(1750)$ & \tiny  &\tiny $\Sigma(1775)$&& \\[-0.5ex]
$70\ [^28]$&**&****&&&\\[-3ex]
                       &\tiny $\Sigma(1620)$  &\tiny $\Sigma(1670)$&&& \\[-0.5ex]
\hline\hline
\end{tabular}
\renewcommand{\arraystretch}{1.0}
\ec
\vspace{-3mm}
\end{table}

\begin{table*}[!htbp]
\caption{%
\label{multip}$N$ and $\Delta$ resonances and their SU(6) multiplet
assignments in the third excitation band. All known resonances in this mass range can be 
assigned to two multiplets. The other six multiplets are empty (from Ref.~\cite{Klempt:2012fy}.  }
\renewcommand{\arraystretch}{1.6}
\begin{center}
\begin{tabular}{||cccccc||}\hline\hline
\hspace{4mm} $J^P=$   & $1/2^-$ & $3/2^-$& $5/2^-$& $7/2^-$& $9/2^-$\\\hline&&&&&\\[-3.6ex]\cline{4-5}
$70\ [^210]$&&&\multicolumn{1}{|c}{-}&\multicolumn{1}{c|}{$\Delta(2200){7/2^-}$}&\\\cline{4-5}&&&&&\\[-3.6ex]\cline{3-6}
$70\ \ [^48]$&&\multicolumn{1}{|c}{$N(2120){3/2^-}$}&\multirow{2}{*}{$N(2060){5/2^-}$}&\multirow{2}{*}{$N(2190){7/2^-}$}&\multicolumn{1}{c||}{$N(2250){9/2^-}$}\\\cline{3-3}\cline{6-6}
$70\ \ [^28]$& &&\multicolumn{2}{|c|}{}&\\\cline{4-5}
&&&&&\\[-3.6ex]\cline{3-5}\cline{2-4}
$56\ [^410]$& \multicolumn{1}{|c}{$\Delta(1900){1/2^-}$}&$\Delta(1940){3/2^-}$&\multicolumn{1}{c|}{$\Delta(1930)5/2^-$}&    & \\\cline{2-4}&&&&&\\[-3.6ex]\cline{2-3}
$56\ \ [^28]$& \multicolumn{1}{|c}{$N(1895){1/2^-}$}&\multicolumn{1}{c|}{$N(1875){3/2^-}$}&&& \\
\hline \hline
           \end{tabular}
\renewcommand{\arraystretch}{1.0}
          \end{center}
\end{table*}
In the third excitation band, a large number of resonances can be expected. In quark
models they fall into one of the following eight representations:
\be
(56,1^-_3), (70, 3^-_3), (56,3^-_3),  (20,3^-_3), \nonumber\\
(70,2^-_3), (70,1^-_3), (70,1^-_3), (20,1^-_3).\phantom{,}
\ee
In the $N$ plus $\Delta$ sector, 30 nucleon and 15 $\Delta$ resonances are expected.
All known states fit into the first two respresentations,
see Table~\ref{multip}.

Particularly interesting are the resonances that can be
assigned to the $(56,1^-_3)$ representation. The 56-plet can be expanded into
\be
56\ =\ ^4\hspace{-1mm}\ 10\ \oplus\ ^2\hspace{-1mm}\ 8\,,
\ee
where the three resonances $\Delta(1900){1/2^-}$, $\Delta(1940){3/2^-}$,
$\Delta(1930)5/2^-$
form a (degenerated) spin-quartet and \linebreak $N(1895){1/2^-}$ and $N(1875){3/2^-}$ a
spin-doublet. In quark models, these states belong to the third excitation band but their masses
are rather compatible with resonances falling into the second excitation band. In these states,
one oscillator is excited to carry one unit of orbital angular momentum, one oscillator carries
one unit of radial excitation. Note that the Roper resonance -- carrying one unit of radial excitation
and belonging to the second excitation shell -- has a smaller mass than $N(1520)3/2^-$ carrying one unit
of orbital angular momentum. Considering the masses of these five resonances above,
we should expect a spin-doublet of negative-parity $\Lambda$ resonances,
and five negative-parity $\Sigma$ states, all only slightly above 2\,GeV. Three one-star candidates are
known, $\Lambda(2000)1/2^-$, $\Lambda(2050)3/2^-$, and $\Sigma(2010)3/2^-$. They could be members
of the $(56,1^-_3)$ representation. We emphasized again that states with identical quantum numbers 
can mix. The mixing angles are, however, predicted to be often small, and experimentally, an assignment
to multiplets seems to be possible.   

Some $N$ and $\Delta$ states can be assigned to the $(70, 3^-_3)$ representation. Expected
are resonances with a total orbital angular momentum of three units. Nucleons in a 70-plet
can carry spin 1/2 or 3/2; $\Delta$'s only spin 1/2. The two pairs of nucleon resonances
with $J^P=5/2^-$ and $7/2^-$ can have intrinsic quark spin $S=1/2$ or $3/2$, respectively, and can be
expected to be separated in mass by about 110\,MeV. So far, they have not been identified
separately. In the remaining multiplets, many more states predicted but no candidates known.

Totally 45 $\Sigma$ resonances are expected in the third excitation band, in a
comparatively narrow mass interval from 2000 to 2400\,MeV.  It seems hopeless
to identify them all. The aim
in a study of $K_L N$ interactions should be to verify that the $N$ and $\Delta$ resonances at about
1900\,MeV with negative parity belong to a 56-plet and have $\Lambda$ and $\Sigma$ partners.
Two $\Lambda$ and five $\Sigma$ states with negative-parity falling into the 2000 to 2100\,MeV mass region 
are to be expected. These are states with one unit of orbital and one unit of radial excitation as dominant
configuration. Also the {\it leading}
resonances with $L=3$ and $S=1/2$ coupling to $J^P=7/2^-$, and $L=3$ and $S=3/2$ coupling to
$9/2^-$ should be identified. In the $\Lambda$ sector, a 4* $\Lambda(2100)7/2^-$ is known that likely
belongs to the SU(3) singlet series (see Fig.~\ref{fig:assignsu3}). The recently suggested 1* 
$\Lambda(2080)5/2^-$~\cite{Sarantsev:2019xxm} 
could be its spin partner. The 1* $\Sigma(2100)7/2^-$ is a bit
low in mass; it could be the strange partner of $N(2190)7/2^-$ or $\Delta(2200)7/2^-$.

\subsection{\boldmath The $\Lambda(1405)$}
The $\Lambda(1405)1/2^-$ resonance was discovered in 1961~\cite{Alston:1961zzd}.
Its spin and parity were first taken from the quark model in which the
$\Lambda(1405)$ and $\Lambda(1520)$ hyperons are interpreted as $qqq$
resonances with a dominant SU(3)-singlet structure~\cite{Isgur:1978xj}. The SU(3)
assignments of $\Lambda(1405)$ and\linebreak $\Lambda(1520)$ as mainly
SU(3) singlet states were confirmed by Tripp {\it et al.}~\cite{Tripp:1969rd} by a
comparison of the
phases at the resonance position of the $K^-p\to \Lambda(1405)\to K N$ and
$K^-p\to \Lambda(1405)\to \pi\Sigma$ transition amplitudes. The preference for
the SU(3)-singlet nature of $\Lambda(1405)$ was statistically significant even though the
data base was meagre. The
spin and parity of $\Lambda(1405)$ were only established in 2014~\cite{Moriya:2014kpv}.

\begin{figure*}[pt]
\begin{center}\includegraphics[width=0.8\textwidth]{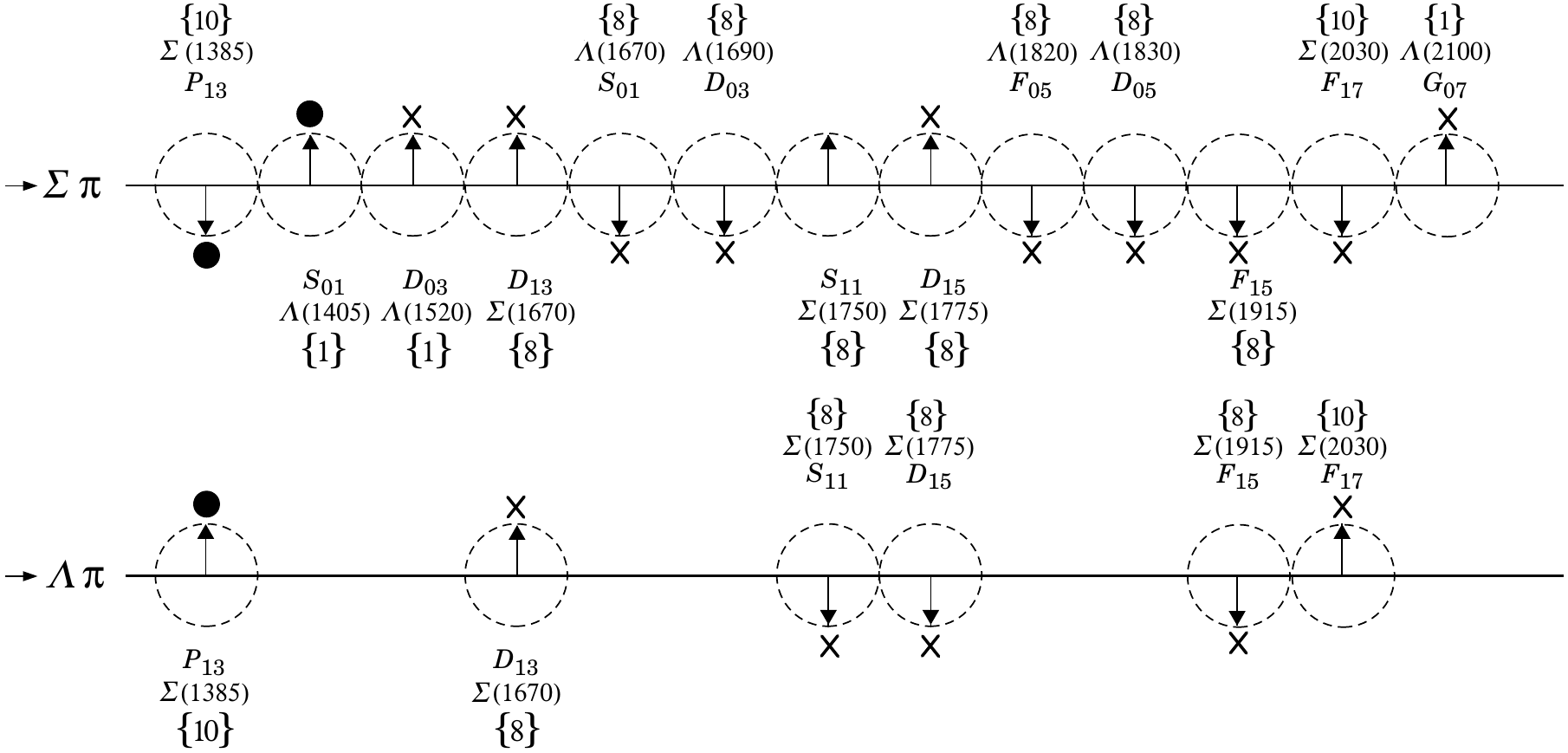}
\vspace{-3mm}\end{center}
\caption{\label{fig:assignsu3}The signs of the imaginary parts of resonating amplitudes in the
$\overline{K} N \rightarrow {\it \Lambda} \pi$ and ${\it \Sigma} \pi$ channels.
The signs of the ${\it \Sigma}(1385)$ and ${\it \Lambda}(1405)$,
marked with a~$\bullet$,
are set by convention, and then the others are determined relative to them.
The signs required by the SU(3) assignments of the resonances are shown
with an arrow, and the experimentally determined signs are shown with an
$\times$. (From Ref.~\cite{Tanabashi:2018oca})
}
\end{figure*}

The SU(3) singlet assignment was challenged by coup\-led-channels calculations
based on chiral SU(3) effective field theories.
Kaiser, Waas and Weise constructed an effective potential from a
chiral Lagrangian, and the $\Lambda(1405)$ emerged as quasi-bound state in the $\bar KN$ and
$\pi\Sigma$ coupled-channel system \cite{Kaiser:1996js}. Oller and Meissner~\cite{Oller:2000fj}
derived the interaction of the SU(3) octet of pseudoscalar mesons and the SU(3) octet of
stable baryons and studied the $S$-wave $\bar K N$ interactions in a relativistic
chiral unitary approach. Two isoscalar resonances at 1379.2\,MeV and at 1433.7\,MeV and
one isovector resonance at 1444.0\,MeV governed the interaction. The two
$\Lambda^*$ poles as well as a third state at 1680\,MeV were interpreted as combinations
of the SU(3) singlet state and the two octet states expected in the expansion $8\otimes8$ into
$1\oplus8_s \oplus 8_a \oplus10\oplus \overline{10}\oplus 27$. The first wider state
(at 1390\,MeV in their analysis) was interpreted as a mainly SU(3)-singlet state, a second and a
third state at 1426\,MeV and 1680\,MeV were interpreted as mainly octet states.
The two expected isovector states were found to be much more sensitive to the details of the
coupled channel approach~\cite{Jido:2003cb}. These results were confirmed in a series of
further studies~\cite{Oset:2001cn,Cieply:2009ea,Cieply:2011nq,Ikeda:2011pi,Ikeda:2012au,Guo:2012vv,Mai:2012dt,Mai:2014xna,Roca:2013av,Roca:2013cca,Miyahara:2018onh,Feijoo:2018den}.
A survey of the literature and a discussion of the different approaches can be found in
Ref.~\cite{Cieply:2016jby}.

The SU(3) structure of a baryon can be deduced from its decays, in particular from the sign
of transition amplitudes at the resonance position. The amplitude for $K^-p\rightarrow
\overline K^0n$ scattering can be decomposed into isospin-0 and -1 elastic scattering amplitudes
$A_0$ and $A_1$ and written as $\pm(A_1 - A_0 ) /2$, where the sign depends on
conventions. It is custom to chose the overall phase so that the amplitude of any
$\Sigma$ at resonance will point ``up'' and any $\Lambda$ at resonance
will point ``down'' (along the negative imaginary axis): The phase at resonance
determines the isospin.

The separation of $\Lambda$ SU(3) singlet and octet states requires a second decay mode,
here $K^-p\to \Lambda(1405)\to\pi\Sigma$. Again, a
convention has to be adopted for some overall phases. We use the convention of
Levi-Setti~\cite{LeviSetti69} that is shown in Fig.~\ref{fig:assignsu3}. The figure
compares experimental results with theoretical
predictions for the signs of several resonances. Since this approach
is not very well known, we first derive the amplitude relations  shown graphically in
Fig.~\ref{fig:assignsu3}.

The decay amplitude of hyperons into a baryon and a meson are governed  by two
SU(3) structure constants, the symmetric ($d_{ijk}$) and the antisymmetric ($f_{ijk}$).
These are tabulated, e.g., in the RPP. Their relative contribution is governed by
the parameter $\alpha$ that depends on the SU(6) classification of the baryon
(see Table~\ref{tab:L-decays}). The corresponding SU(6) coupling constants can be found
in Refs.~\cite{Guzey:2005rx,Guzey:2005vz}) and are listed in Table~\ref{tab:L-decays}.
The production cancels in the comparison, and the relative sign of the amplitudes can be
used to determine the SU(3) structure of a hyperon. The relative signs are listed
in the last line in Table~\ref{tab:L-decays}. Mixing of the $^28[70]$ component into
the  $^21[70]$ wave function could reverse the sign from $+1$ to $-1$, which would
make $\Lambda(1405)$ appear as ``mainly'' octet state. 

\begin{table}[pt]
\caption{\label{tab:L-decays}SU(3) coupling constants for hyperon decays and the SU(6) predictions
for the coefficient $\alpha$ in decays of octet hyperons.
}
\renewcommand{\arraystretch}{1.35}
\bc
\begin{tabular}{cccc}
\hline\hline
Decay mode                &$8\to 8+8$ & $1\to 8+8$           & \\\hline
$\Lambda\to N\bar K$  &$\sqrt{\frac23}(2\alpha +1)A_8$& $\frac12 A_1$       & \\
$\Lambda\to \Sigma\pi$&$2(\alpha-1)A_8$             &$\sqrt{\frac32}A_1$& \\\hline
&$^28[56]$& $^28[70]$ & $^48[70]$ \\
$\alpha$&$\frac25$ & $\frac58$ & $-\frac{1}{2}$\\
\hline
\end{tabular}

\renewcommand{\arraystretch}{1.6}
\begin{tabular}{ccccc}
&$^21[70]$&$^28[56]$& $^28[70]$ & $^48[70]$ \\\hline
$\frac{A(\Lambda\to N\bar K)}{A(\Lambda\to \Sigma\pi)}$&$\sqrt{\frac16}$&  $-\sqrt{\frac32}$&   $-\sqrt6$ &    0   \\
Sign &$+$&$-$&$-$&\\
\hline\hline
\end{tabular}
\vspace{-4mm}
\ec
\end{table}

The BnGa collaboration analyzed the CLAS data on the three charge states in
$\gamma p\to K^+ (\Sigma\pi)$ \cite{Moriya:2013hwg}, combined with data on
the total cross sections for $K^-p$ induced reactions: $K^-p\to K^-p$,
$K^-p\to\bar K^0n$, $K^-p\to\pi^0\Lambda$, $K^-p\to\pi^+\Sigma^-$, $K^-p\to\pi^0\Sigma^0$,
$K^-p\to\pi^-\Sigma^+$~\cite{Humphrey:1962zz,Watson:1963zz,Sakitt:1965kh,Ciborowski:1982et}, elastic and inelastic $K^-p$ scattering~\cite{Mast:1975pv}, the low-energy BNL data on
$K^-p\to\pi^0\pi^0 \Lambda (\Sigma^0)$ \cite{Prakhov:2004ri,Prakhov:2004an},
bubble chamber data on $K^-p\to\pi^-\pi^+\pi^{\pm}\Sigma^{\mp}$
\cite{Hemingway:1984pz}, $K^-p$ annihilation frequencies at
rest~\cite{Tovee:1971ga,Nowak:1978au}, and the recent experimental results on the energy
shift and width of kaonic hydrogen atoms which constrain the $K^-p$ $S$-wave scattering length
\cite{Bazzi:2011zj,Bazzi:2012eq}. Very important are the data on $K^-p\to \pi\Sigma$
\cite{Armenteros:1970eg} which constrain the SU(3) structure of $\Lambda(1405)$.
In the preferred solution, the BnGa partial-wave analysis~\cite{Anisovich:2020tbd} required only one isoscalar
resonance with a pole at
[(1421\er 3)-i((23\er3)]\,MeV. The pole can be identified with the $\Lambda(1405)$ at a slightly 
higher mass compared to the
nominal mass. The isovector interactions were described by two resonances, one below, one
above the considered mass range (1300 - 1500\,MeV).  The SU(3) structure was determined to
be consistent with a singlet but not with an octet state. There was, however, a second solution with
a description of the data with similar quality. This second solution was compatible with a second
broader isoscalar resonance with a fixed mass at 1380\,MeV. In this solution, the  $\Lambda(1405)$ changed its SU(3)
structure from being dominant SU(3) singlet to dominant SU(3) octet.  Obviously,  the $\Lambda(1405)$  
SU(3) structure cannot be determined in a model-independent way from existing
$K^-p$ scattering alone, even when the CLAS data
on photo-induced data on $\Lambda(1405)$ production are included in the analysis.

The $K^-p$ threshold is at 1432\,MeV, considerably above the nominal $\Lambda(1405)$
mass. At present, data on differential cross sections for $K^-p\to \Lambda(1405)\to K N$ exist only above
1470\,MeV, those for  $K^-p\to \Lambda(1405)\to \pi\Sigma$ only above 1530\,MeV.
It will be important to repeat the BnGa analysis with data on $K^-p$ scattering
covering a mass range starting from close to the threshold to about 1540\,MeV.

\begin{table}[pt]
\caption{\label{1670}The signs of the SU(6) amplitudes for
$\Sigma^+(1670)3/2^-\to\pi^+\Lambda(1405); \ \Lambda(1405)\to\Sigma^\pm\pi^\mp$ and
$ \Sigma^+(1670)3/2^-\to\pi^+\Lambda(1405); \ \Lambda(1405)\to\Sigma^\pm\pi^\mp$}
\bc
\renewcommand{\arraystretch}{1.4}
\begin{tabular}{ccccc}
\hline\hline
\multicolumn{3}{r}{$\Lambda(1405)$ SU(3) structure:}       & 1 & 8\\\hline
$\Sigma^+(1670)3/2^-$&$\to$&$\Lambda(1405)\pi^+$ &+&+\\
                                  &        &$\hookrightarrow\Sigma^\pm\pi^\mp$ & +&-\\
\multicolumn{3}{r}{Sign of transition amplitude at pole:}& +&-\\\hline
$\Sigma^+(1670)3/2^-$&$\to$&$\Sigma^0(1385)\pi^+$ &+&+\\
                                 &         &$\hookrightarrow\Sigma^\pm\pi^\mp$&+&+\\
\multicolumn{3}{r}{Sign of transition amplitude at pole:}& +&+\\
\hline\hline
\end{tabular}
\vspace{-3mm}\ec
\end{table}

In the reaction $K^-p\to\pi^-\pi^+$ $\pi^{\pm}\Sigma^{\mp}$ studied
in~\cite{Hemingway:1984pz}, the full $\Lambda(1405)$ line shape can be investigated.
In this reaction, the SU(3) assignment follows from the correlation in the production and decay dynamics.
The derivation relies on approximate SU(6) symmetry in baryon decays.
We consider the two decay sequences
\begin{subequations}
 \label{LS}
\begin{align}
\Sigma^+(1670)3/2^-\to\pi^+\Lambda(1405); \ \Lambda(1405)\to\Sigma^\pm\pi^\mp \hspace{1mm}\label{L1}\\
\Sigma^+(1670)3/2^-\to\pi^+\Sigma(1385); \ \Sigma(1385)\to\Sigma^\pm\pi^\mp \label{S1}
\end{align}
\end{subequations}
that are shown to contribute to this reaction~\cite{Anisovich:2020tbd}.

The SU(6) amplitude for reaction~(\ref{L1}) depends on the
SU(3) structure of $\Lambda(1405)$ and on the primary $\Sigma^+(1670)$ $3/2^-$
(see Table~\ref{1670}).
The sign of this amplitude is given by the product of the signs for
$\Sigma^+(1670)3/2^-\to\Sigma\pi$ and $\Lambda^+(1405)1/2^-\to\Sigma\pi$. The
$\Sigma^+(1670)3/2^-$ belongs dominantly to a spin-1/2 SU(3) octet in the SU(6) 70-plet;
$\alpha=5/8$. The sign of the SU(6) amplitude for $\Sigma^+(1670)3/2^-$ $\to\Sigma\pi$ is
given by $2\sqrt 2\cdot\alpha$, hence $+1$; the sign for the $\Lambda^+(1405)\to\Sigma\pi$
transition depends on the SU(3) structure of  $\Lambda^+(1405)$: if it is an octet with spin-1/2
in the SU(6) 70-plet, it is given by $2(\alpha -1)$ with $\alpha=5/8$, hence negative. If it
is a singlet, it is $\sqrt{6/4}$ and positive.  The sign of the transition amplitudes for
reactions~(\ref{L1}) and ~(\ref{S1}) are the same
when $\Lambda(1405)$ is an octet, they are different when $\Lambda(1405)$ is an octet.

\section{\label{3rd}The positive-parity states in the second excitation band}
\subsection{Missing resonances}
The second excitation band contains a number of representations:
\be
(56,0_0 ^+); (70,0_2 ^+); (56,0_2 ^+); (70,0_2 ^+); (20,1_2 ^+)\,.
\ee
In total, there are 8 $\Delta$ and 8 $\Omega$ resonances expected in the 2$^{\rm nd}$ excitation shell,
13 nucleon resonances, 19 $\Lambda$ resonances, and 21 $\Sigma$ and 21 $\Xi$ resonances. 
The Particle Data Group classifies baryon resonances with a star rating; 3* and 4* resonances are considered
to be established, 1* and 2* resonances not.  Table~\ref{tab:mult2} gives the number of predicted states 
and compares this number with the number of established and the number of 1* or 2* states.

\begin{table}[ph]
\caption{\label{tab:mult2}Number of expected and observed resonances that can be assigned to the 2$^{\rm nd}$ excitation shell
for $J^P=1/2^+, .. , 7/2^+$. The first number gives the expected number of resonances, followed by the number of observed
resonances with 3* and 4*, 1* and 2* (in parentheses).
}
%\scriptsize
\renewcommand{\arraystretch}{1.35}
\bc
\begin{tabular}{ccccccc}
\hline\hline
&&$1/2^+$&$3/2^+$&$5/2^+$&$7/2^+$&Sum\\\hline
seen&$N$ &\hspace{-2mm}4 (4,0)&5 (3,1)&3 (1,2)&1 (1,0)&13 (9,3)\\
seen&$\Delta$ &\hspace{-2mm}2 (1,1)&3 (2,0) &2 (1,0)&1 (1,0)& 8 (5,1)\\
seen&$\Lambda$ &6 (2,1)&7 (1,1)&5 (2,0)&1 (0,1)&19 (5,3)\\
seen&$\Sigma$ &6 (1,1)&8 (0,4)&5 (1,1)&2 (1,0)&21 (3,6)\\
seen&$\Xi$ &6 (0,0)&8 (0,0)&5 (0,0)&2 (0,0)&21 (0,0)\\
seen&$\Omega$&2 (0,0)&3 (0,0)&2 (0,0)&1 (0,0)&8 (0,0)\\
\hline\hline\vspace{-8mm}
\end{tabular}
\ec
\end{table}
 In the nucleon spectrum, thirteen states  are expected in the second excitation level. Nine states 
are established, three states need further confirmation, one state is missing. The number of
$J^P=1/2^+$ states seems complete; yet the state with highest mass, $N(2100)1/2^+$, may already
belong to the fourth excitation shell. (It could be low in mass like the Roper resonance in the second excitation shell,
see Ref.~\cite{Loring:2001ky}.)
Then, one state would be missing. For $J^P=3/2^+$, one state is missing. 
Below we will discuss the reasons
why we might expect not to observe the two nucleon states (with $J^P=1/2^+$ and $3/2^+$) in the
20-plet. In the $\Delta$ spectrum, one state with $J^P=3/2^+$, one 
with $J^P=5/2^+$ are missing, one further states with $J^P=1/2^+$ is seen with little evidence only. 
 The situation is much worse
in for $\Lambda$ and $\Sigma$ hyperons: only 17 of 42 states are seen, only 8 of them are established.
No $\Xi$ resonance or $\Omega$ with known spin-parity that might belong to the second excitation shell
is listed in the RPP.

In $K_Lp$ scattering experiments, $\Sigma$ resonances can be searched for in formation. $\Lambda$
resonances are formed only by scattering off neutrons.
The reactions
\begin{subequations}
 \label{binaries}
\begin{align}
K_L p&\to\Sigma^{+*}\to K_Sp &\
&\\
K_L p&\to\Sigma^{+*}\to  \pi^+\Lambda; \
\pi^+\Sigma^0; \  \pi^0\Sigma^+
\end{align}
\end{subequations}
and there analysis, reconstruction and partial wave analysis, were described in detail in Ref.~\cite{Amaryan:2017ldw}.
Here, we refrain from further discussions.

\subsection{On the sideline: \boldmath$\Delta^{++}$ excitations}
A $K_L$ beamline can also be used to study $\Delta$ excitations in the reaction
\be
\label{d++}
K_L p\to K^-\Delta^{++}\
\ee
The advantage is similar to $\pi^+p$ scattering: only $\Delta$ and no $N$ excitations can be
produced. Admittedly, $\pi^+p$ scattering as formation experiment is superior.

\subsection{Search for the states in the SU(3) 20-plet}

The baryonic spatial wave can be constructed from the two degrees of freedom of a three-particle
system (neglecting the cms motion). In the three-particle system, two quarks can oscillate (the $\rho$ oscillator)
or two quarks can oscillate against the third quark (the $\lambda$ oscillator).  With these oscillators spatial wave
functions can be formed that are symmetric with respect to (w.r.t.) the exchange of any pair of quarks
  
 {\footnotesize
\begin{equation}
\label{S}
S= \frac{1}{\sqrt{2}}\bigl\lbrace
    \left[\phi_{0s}(\vec \rho)\times\phi_{0d}(\vec \lambda)\right]
    +
    \left[\phi_{0d}(\vec \rho)\times\phi_{0s}(\vec \lambda)\right]\bigr\rbrace^{(L=2)},
\end{equation}}
or they can have mixed symmetry
  {\footnotesize
\begin{subequations}
\begin{align}
\label{MS}
\mathcal{M_S}
    &=&\hspace{-3mm}\frac{1}{\sqrt{2}}\bigl\lbrace
    \left[\phi_{0s}(\vec \rho)\times\phi_{0d}(\vec \lambda)\right]-\left[\phi_{0d}(\vec \rho)\times\phi_{0s}(\vec \lambda)\right]\bigr\rbrace^{(L=2)} \\
%\end{eqnarray}\begin{eqnarray}
\label{MA}\mathcal{M_A}&=&
    \left[\phi_{0p}(\vec \rho)\times\phi_{0p}(\vec \lambda)\right]^{(L=2)}\,,
\end{align}
\end{subequations}
}
with one part that is symmetric in the $\rho$ and antisymmetric in the $\lambda$ oscillator
and one part antisymmetric in the $\rho$ and symmetric in the $\lambda$ oscillator.
Both parts are required in the full wave function. The part $\mathcal{M_A}$ describes a
component in which the $\rho$ and the $\lambda$ oscillator are both excited simultaneously.

Finally, the spatial wave function can be antisymmetric w.r.t. the exchange
of any quark pair:
{\footnotesize\begin{eqnarray}
\label{A}
\mathcal{A} & = &
    \left[\phi_{0p}(\vec \rho)\times\phi_{0p}(\vec \lambda)\right]^{(L=1)}\,.
\end{eqnarray}}
 The multiplets 70, 56, and 20 arise from the combination of the three light quarks $u,d,s$
having spin 1/2:
\be
6\otimes6\otimes6=56_S\oplus70_M+20_A
\ee

The spin-flavor wave functions can be symmetric ($S$) or antisymmetric ($A$) or can be of
mixed symmetry. In the ($M$).  Since the total spin-flavor-spatial wave function needs
to be symmetric, the spatial wave function has to carry the same symmetry. Hence the
antisymmetric wave function~(\ref{A}) is combined with the spin-flavor wave function of
the 20-plet.

\begin{figure}[pt]
\begin{center}\includegraphics[width=0.48\textwidth,height=0.26\textwidth]{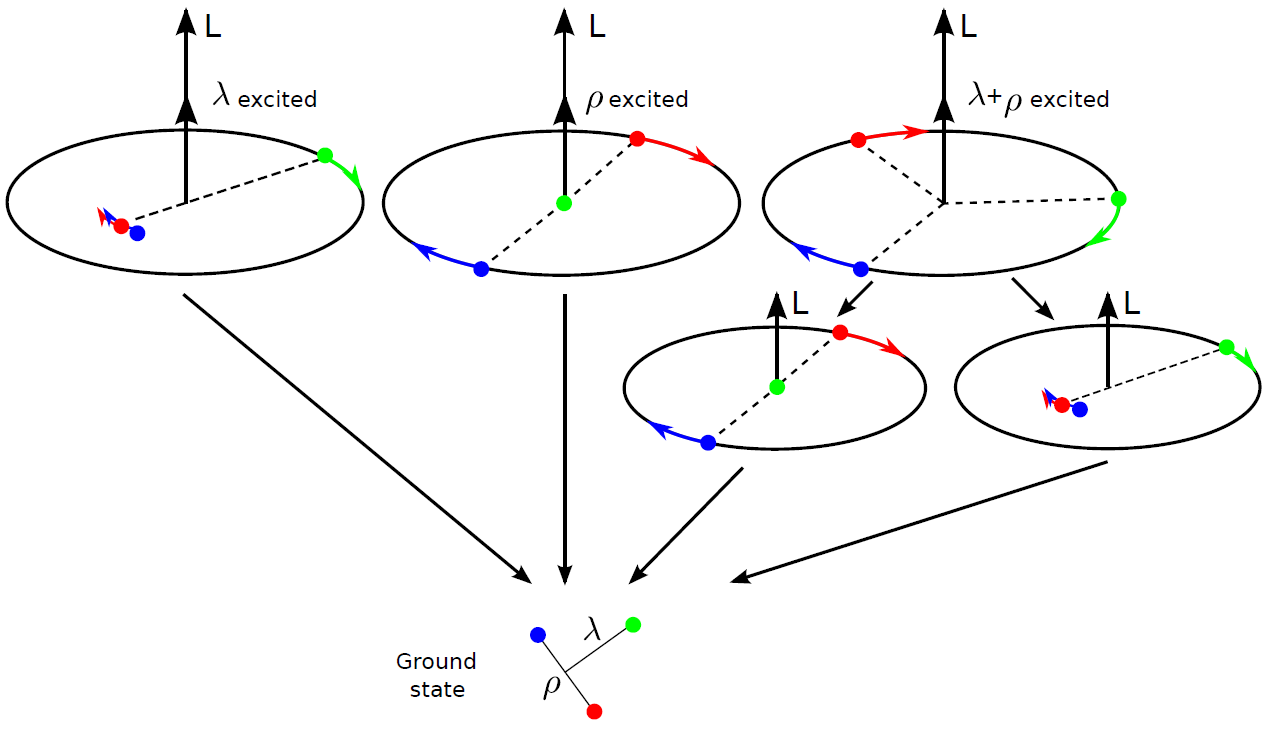}
\vspace{-3mm}
\end{center}
\caption{\label{fig:rho-lambda}(Color online) Classical orbits of nucleon excitations with L = 2
(upper row) and L = 1 (lower row). The first two pictures in
both rows show excitations of the $\rho$ and $\lambda$ oscillators, in the
third picture in the first row both, $\rho$ and $\lambda$ are excited~\cite{Thiel:2015kuc}.\vspace{-3mm}
}
\end{figure}

In an analysis of data on the reaction $\gamma p\to\pi^0\pi^0p$~\cite{Sokhoyan:2015fra}
it was shown that the dominant decay modes of resonances like the quartet of $\Delta$ resonances, $\Delta(1910)1/2^+$,
$\Delta(1920)3/2^+$, $\Delta(1905)5/2^+$, $\Delta(1950)7/2^+$, having symmetric wave
functions of type~(\ref{S}), are decays into two ground-state hadrons like $N\pi$ or $\Delta\pi$.
Nucleon resonances in the 70-plet with a mixed symmetry, (\ref{MS}) and (\ref{MA}), have sizable
branching ratios into final states in which one of the decay product carries orbital excitation like
$N(1520)\pi$ or $Nf_{{\pi\pi}_{\rm S-wave}}$. This observation was interpreted as evidence for
the three-body nature of nucleon excitations~\cite{Thiel:2015kuc}. The situation is depicted in
Fig.~\ref{fig:rho-lambda}. When the $\lambda$ or $\rho$ oscillator is excited, it can de-excite
into the ground state. When both oscillators are excited, then first one oscillator de-excites into an
intermediate excited states while the other oscillator remains in an excited state. A second step, a
cascade, is required to reach the final state.

This scenario forbids (or suppresses) a direct excitation of resonances that have no symmetric component.
Reversing the argument, it forbids excitations from the ground state into resonances having a wave function
of type~(\ref{A}). The antisymmetry of the orbital wave function needs to be combined with
an antisymmetric spin-flavor wave function. These belong to the 20-plet representations.
The 20-plet can be expanded into
\be
20\ =\ ^28\ \oplus\ ^4 1\,.
\ee
So far, no member of a 20-plet has ever been identified. In the case of nucleon resonances,
this is rather difficult: there are several possibilities to realize nucleon excitations with internal
total quark spin 1/2. But the discovery of a $\Lambda$ state that decays mostly via a cascade
process would provide strong evidence that a member of a 20-plet has been identified. The
observation of a series of states with $J^P=1/2^+, 3/2^+, 5/2^+$  decaying via cascades
would strengthen the conjecture.

\begin{figure*}[h!]
\vspace*{-15mm}
\bc
\begin{minipage}[c]{0.27\textwidth}
\large
\vskip 25mm
\renewcommand{\arraystretch}{1.5}
\hspace{-12mm}\begin{tabular}{l}
\hspace*{50mm}$\rm  uud d\bar s$\\
\\
\hspace*{17mm} $\rm  udd(\frac{1}{\sqrt{3}}\, d\bar{d} + \sqrt{\frac{2}{3}}\,  s \bar{s})$\\
\\
\hspace*{8mm}$\rm  dd{ s}(\frac{1}{\sqrt{3}}\, { s\bar{s}} +
\sqrt{\frac{2}{3}}\, d \bar{d})$\\
\\
\hspace*{20mm}$\rm  dd{ ss} \bar{u}$\\
\end{tabular}\vspace{10mm}
\renewcommand{\arraystretch}{1}
\end{minipage}
\begin{minipage}[c]{0.45\textwidth}
\begin{center}
\setlength{\unitlength}{0.6mm}% antidecuplet
\begin{picture}(150.00,90.00)
\put(10.00,20.00){ \vector(1,0){70.00}}
\put(45.00,-10.00){\vector(0,1){90.00}}
\put(82.50,20.00){\makebox(5.00,5.00){$I_3$}}
\put(47.50,82.00){\makebox(5.00,5.00){ Antidecuplet }}
\put(07.50,-5.00){\circle*{2.00}}
\put(5.8,-5.00){  \circle*{4.80}}
\put(32.50,-5.00){\circle*{2.00}}
\put(57.50,-5.00){\circle*{2.00}}
\put(80.50,-5.00){  \circle*{4.80}}
\put(45.00,20.00){\circle*{2.00}}
\put(70.00,20.00){\circle*{2.00}}
\put(20.00,20.00){\circle*{2.00}}
\put(32.50,45.00){\circle*{2.00}}
\put(57.50,45.00){\circle*{2.00}}
\put(43.50,70.00){  \circle*{4.80}}
\put(7.50,-5.00){\line(1,0){75.00}}
\put(82.50,-5.00){\line(-1,2){37.50}}
\put(7.50,-5.00){\line(1,2){37.50}}
\put(76.00,67.50){\makebox(15.00,5.00)[r]{$\mathbf \Theta^+(1530)$\ \  S=+1}}
\put(17.00,42.50){\makebox(13.25,5.00)[r]{$  N^0{(1710)}$}}
\put(90.00,42.50){\makebox(13.25,5.00)[r]{$  N^+{(1710)}$\ \  S=+0}}
\put(-11.00,12.50){\makebox(12.50,5.00)[l]{$  \Sigma^-(1890)$}}
\put(35.00,12.50){\makebox(12.50,5.00)[l]{$  \Sigma^0(1890)$}}
\put(80.00,12.50){\makebox(12.50,5.00)[l]{$  \Sigma^+(1890)$\quad  S=-1}}
\put(-11.00,-15.00){\makebox(12.50,5.00)[l]{$  \bf \Xi^{--}(2070)$}}
\put(22.00,-15.00){\makebox(12.50,5.00)[l]{$  \Xi^{-}(2070)$}}
\put(55.00,-15.00){\makebox(12.50,5.00)[l]{$  \Xi^{0}(2070)$}}
\put(85.50,-15.00){\makebox(12.50,5.00)[l]{$\mathbf\Xi^+(2070)$\quad  S=-2}}
\end{picture}
\end{center}
\end{minipage}
\hspace*{-15mm}
\begin{minipage}[c]{0.27\textwidth}
\large
\vskip 25mm
\renewcommand{\arraystretch}{1.5}
\begin{tabular}{l}
\hspace*{-5mm}$\rm  uud d\bar s$ \\
\\
 $\rm  uud(\frac{1}{\sqrt{3}}\, d\bar{d} + \sqrt{\frac{2}{3}}\,  s \bar{s})$\\
\\
\hspace*{10mm}$\rm  uu{ s}(\frac{1}{\sqrt{3}}\, { s\bar{s}} +
\sqrt{\frac{2}{3}}\, d \bar{d})$\\
\\
\hspace*{20mm}$\rm  uu{ ss} \bar{d}$\\
\end{tabular}\vspace{10mm}
\renewcommand{\arraystretch}{1}
\end{minipage}
\caption{The antidecuplet and its quark model decomposition.
The antidecuplet predicted by the chiral soliton model
describes nucleons in terms of the pion field and not
by the number of quarks~\protect\cite{Diakonov:1997mm}.
The three corner-states are incompatible with a $qqq$ assignment.}
\label{anti10}
\ec
\vspace*{-5mm}
\end{figure*}
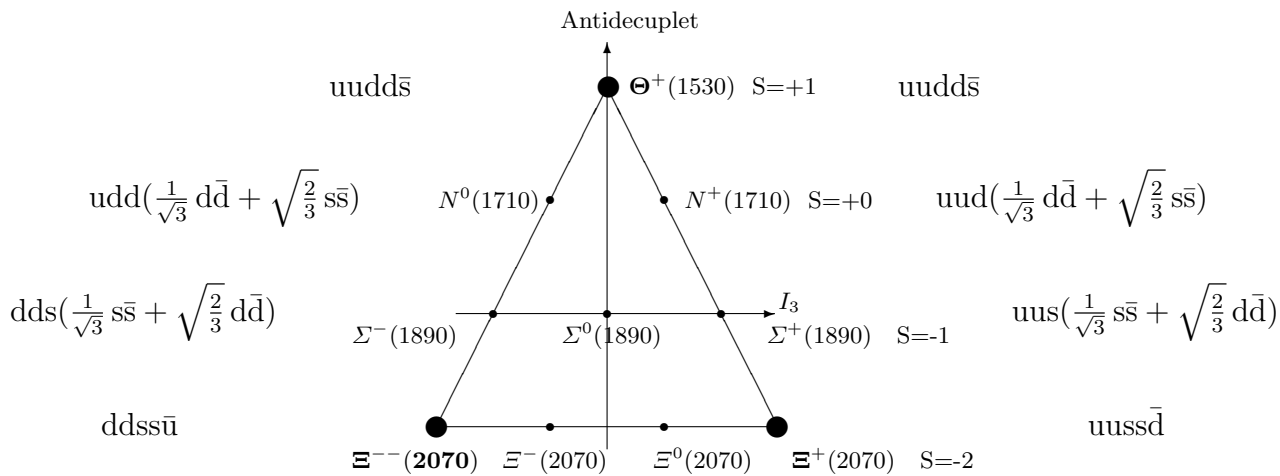

In Ref.~\cite{Loring:2001ky} the three states are predicted to have masses of 2099\,MeV;
2176\,MeV; 2150\,MeV. We suggest to search first for the member of the 20-plet
with $J^P=3/2^+$ in the reaction
\be
K_L\,p\to\,\pi^+\Lambda_{20}\,,\quad\Lambda_{20}\to\Lambda(1520)\eta\   {\rm or} \  \Lambda(1670)\eta\,.
\ee
This is an $S$ wave decay to an intermediate state with orbital angular momentum excitation.
The first decay mode has the disadvantage that $\Lambda(1520)$ is dominantly a SU(3) singlet,
$\eta$ dominantly SU(3) octet but the mixing angles deviate significantly from pure SU(3) eigenstates.
The second mode might be forbidden kinematically if the mass of the expected resonance is low.
With $L=2$ between $\eta$ and exciated hyperon, also the  states with $J^P=1/2^+$ and $5/2^+$ could be observed.
Note that $\Lambda$ excitations with a total quark spin $S=3/2$ exist only in the SU(6) 20-plet.

\subsection{Pentaquark search}

The concept of a nucleon composed of three constituent quarks is certainly oversimplified, and the hadronic properties
of nucleons cannot be understood or, at least, are not understood in terms of quarks and their interactions. Skyrme
studied the pion field and discovered that by adding a non--linear ``$\sigma$ term'' to the pion field equation, stable
solutions can result~\cite{Skyrme:1961vq}. These solutions have half integer spin and a winding number
identified by Witten~\cite{Witten:1983tw} as the baryon number. These stable solutions of the pion field
equation are called soliton solutions.

The chiral soliton model predicts the existence of a full antidecuplet
of states~\cite{Chemtob:1985ar,Walliser:1992vx} with quantum numbers
$J^P=1/2^+$. The antidecuplet is shown in Fig.~\ref{anti10}; the states are called
pentaquarks~\cite{Diakonov:1997mm}. 
Note that the three corner states have quantum numbers which cannot
be constructed out of three quarks. In the minimum quark model configuration,
the flavor wave function of the state with positive strangeness 
is given by $\Theta^+ = uud d\bar s$. The strange quark fraction
increases from 1 to 2 units in steps of 1/3 additional $s$ quark.
The masses of the pentaquark states were predicted in Ref.~\cite{Diakonov:1997mm}.
The increase in mass per unit of strangeness is is 540\,MeV, instead of the 120\,MeV that are derived
when the $\rho$ or $\omega$ mass is compared to the K$^*$ mass.
The splitting is related to the so--called $\sigma_{\pi N}$ term
in low--energy $\pi$\,N scattering. Its precise value is difficult
to determine and has undergone a major revision~\cite{Diakonov:2003jj}.

Pentaquarks were highly discussed when the so-called  $\Theta^+$ was observed in
different experiments~\cite{Nakano:2003qx,Barmin:2003vv,Barth:2003es,Stepanyan:2003qr}.
It has positive strangeness $S=+1$,
its flavor wave function has a minimal quark content $uud d\bar s$. However,
in a series of precision experiments, the evidence for pentaquarks has faded away
(see, e.g., Ref.~\cite{Dzierba:2004db,Danilov:2007bp,Liu:2014yva}) even though some evidence remains that
a narrow state with $J^P=1/2^+$ at 1720\,MeV might exist \cite{Kuznetsov:2014aka,Kuznetsov:2015nla,Gridnev:2016dba}.
High-precision experiments are mandatory to settle this important issue.
Particularly convincing would be, of course, the discovery or confirmation of one o the
states having quantum numbers that are incompatible with a $qqq$ interpretation.

Attractive and easily accessible is the $\Theta^+$. It is best searched for in the
reaction
\be
\label{theta-elastic}
K_L p\to K^+ n\,.
\ee
The reaction does not receive contributions from $\Sigma$ resonances, nor from Pomeron
exchange nor from the exchange of $f_0 / f_2$ mesons. In this paper, 
we concentrate on inelastic scattering processes and do not expand on
reaction~(\ref{theta-elastic}).

\begin{figure}[pt]
\bc
\begin{tabular}{cc}
\includegraphics[width=0.23\textwidth,height=0.15\textwidth]{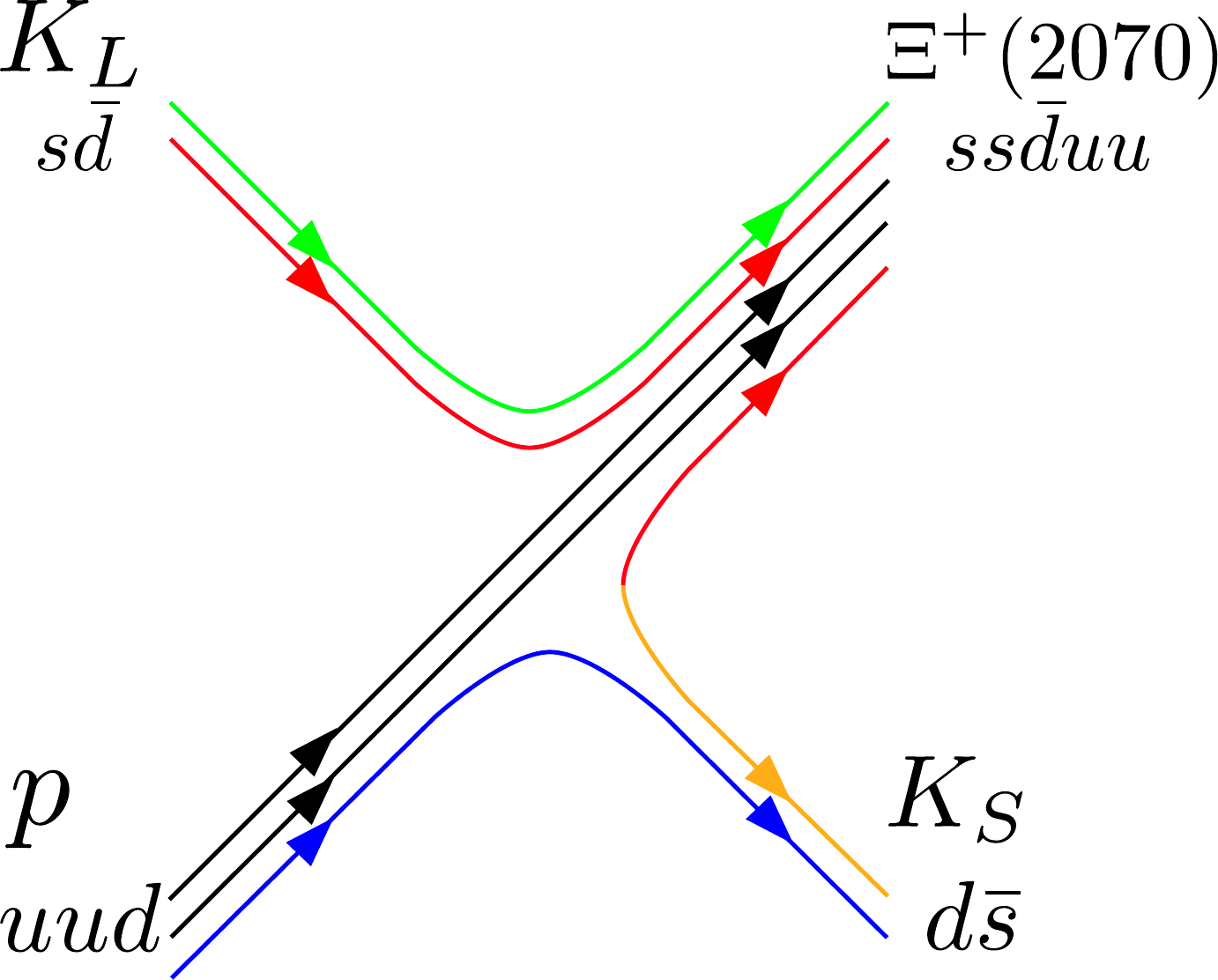}&\hspace{-4mm}
\includegraphics[width=0.23\textwidth]{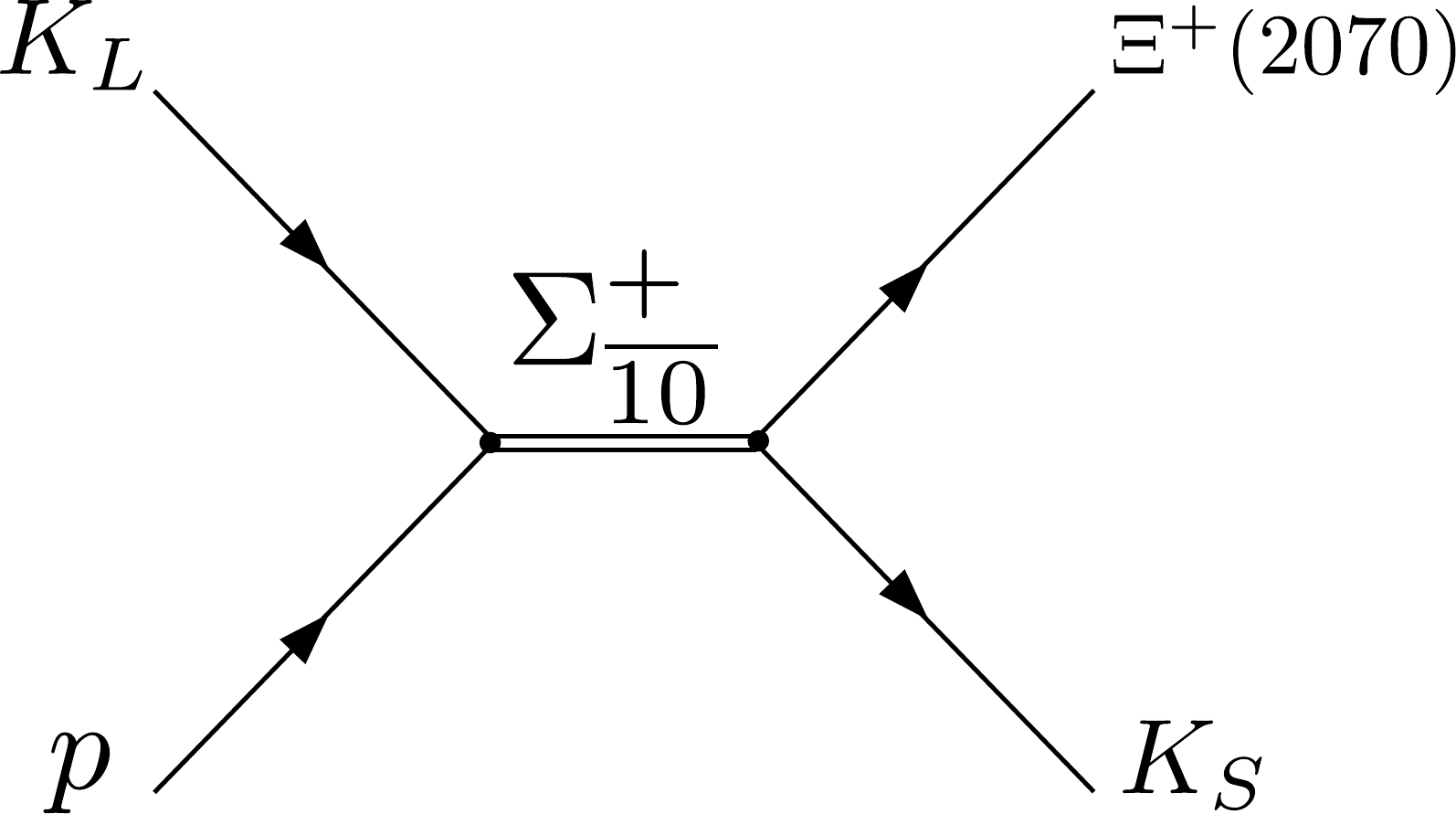}
\end{tabular}
\ec
\caption{\label{pic:2070}(Color online)Left: Quark flow diagram for the reaction $K_L p\to K_S \Xi^+(2070)$.
$s$-quarks in red, $\bar s$ in orange, $d$-quarks in blue, $\bar d$ in green, $u$-quarks in black.
Right: Hadron representation of the scattering process.\vspace{-2mm}
}
\end{figure}

Particularly interesting is the search for a member of the quartet of $\Xi$ pentaquarks.
The minimal quark content of the $\Xi^+(2070)$ is $uuss\bar d$. It can be produced in the $K_L p$ induced
reaction
\be
\label{2070}
K_L p\to K_S \Xi^+(2070)
\ee
At the first moment, the reaction looks like an elastic scattering process. However, the reaction~(\ref{2070}) is more
complicated. The minimal quark flow is depicted in Fig.~\ref{pic:2070}. The process can be described as formation of a 
$\Sigma^+$ state belonging to the antidecuplet.

Evidence for an isospin partner of $\Xi^+(2070)$ with $S = -2$, $Q = -2$ was reported \cite{Alt:2003vb}
studying proton proton collisions at the CERN SPS. Its mass of (1862\er 2)\,MeV was a bit low when compared to
the prediction~\cite{Diakonov:1997mm}. The state was not confirmed in later experiments~\cite{Dzierba:2004db}.

The $\Xi^+(2070)$ is best searched for in its decay into $\Xi^0\pi^+$, predicted with 30\%
branching ratio, followed by the decay $\Xi^0\to \Lambda\pi^0$
($\sim 100$\%). Thus the reaction
\be
K_L p \to K_S\pi^+\pi^0\Lambda\qquad \Lambda \to p\pi^-; K_S\to \pi^+\pi^-
\ee
needs to be studied. The $K^0$ mass and momentum can be reconstructed from the $\pi^+\pi^-$
pair. With a known $K_L$ momentum, the $\Xi^+(2070)$ mass and momentum can be determined.
Then, using the $\pi^+$ four-vector, the $\Xi^0$ mass and momentum can be deduced.
The $\Lambda$ mass and momentum can be deduced from its decay particles; the crossing of
the $\Xi^0$ and $\Lambda$ trajectories defines the decay point of the $\Xi^0$.
The $\Xi^0$ has a mean free path $c\tau=8.71$\,cm. Thus, the reaction
chain will be reconstructed with very little background. An alternative attractive decay mode is given by
$\Xi(2070)\to K ^{*+}\Sigma^0$. The threshold for this decay mode is 2084\,MeV.

The non-strange and strange partners in the anti-decu\-plet suffer from the difficulty
that their identification as members of the anti-decuplet is model-dependent. Evidence
for the possible existence of two narrow states at 1685 and 1720\,MeV has been
reported \cite{Kuznetsov:2014aka,Kuznetsov:2015nla,Gridnev:2016dba}. The peak at 1685\,MeV is discussed
extensively in the literature, see, e.g.,Refs.~\cite{Anisovich:2015tla,Werthmuller:2015owc,%
Witthauer:2017get,Anisovich:2017xqg,Kuznetsov:2017xgu,Kuznetsov:2017qmo,Kuznetsov:2018dcd}.
It seems to belong to the $J^P=1/2^-$ partial wave and to be unrelated to
pentaquark spectroscopy. The structure at 1720\,MeV certainly requires further
investigations but we do not see a particular advantage to use a $K_L$ beam.

There is a triplet of $\Sigma$ states in the antidecuplet. It is
predicted to mix with its $nns$-partners. In Ref.~\cite{Goeke:2009ae}
the mass of the additional mainly-$\bar{10}$ state is calculated to fall into the range
$1795<M_{\bar{10}}<1830$\,MeV; its main decay modes with estimated branching ratios
of nearly 60\% (16\%) are $\bar KN$ ($\pi\Lambda$). The $\Sigma^+$ decuplet state
can be searched for in a formation experiment. The main difficulty is to identify
it against the expected $nns$ states. Quark models, e.g. the Isgur quark model,
predicts six $J^P=1/2^+$ states in the second excitation band at 1720, 1915, 1970,
2005, 2030, 2105\,MeV. Given the uncertainties with the calculation of Roper-like
states in the quark model and the uncertainty of the predictions using the chiral soliton
model, there is certainly a significant model-dependence in any attempt to assign
a specific state with non-exotic quantum numbers to the antidecuplet.
\begin{figure}[pt]
\bc
\includegraphics[width=0.4\textwidth,height=0.34\textwidth]{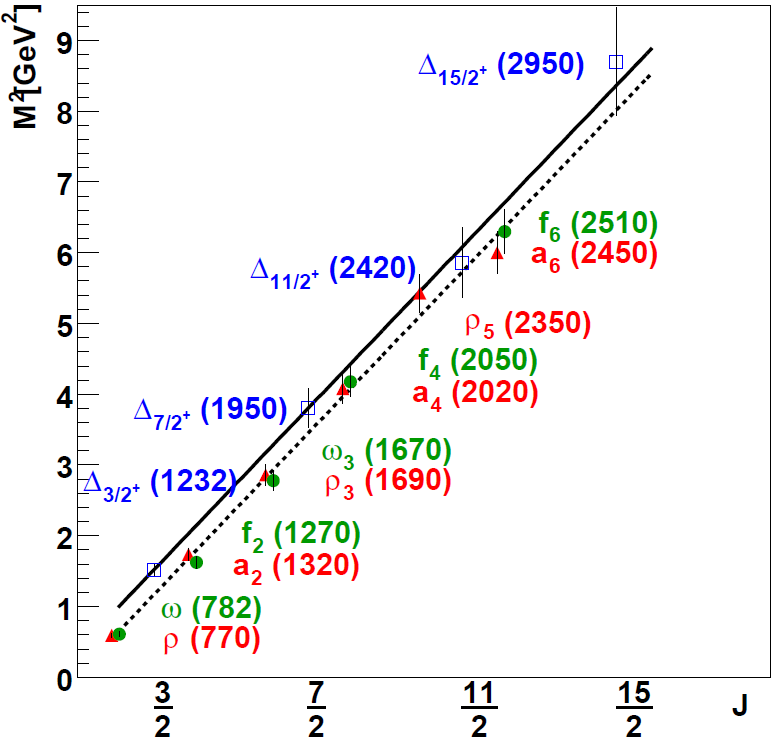}\\
\ec
\caption{\label{pic:regge}The Regge trajectories $M^2$ versus $J$ for mesons and
$\Delta$ baryons have the same slope. This observation suggests for stretched states with $J=L+S$
a string excitation between a quark and a diquark in baryons (from Ref.~\cite{Klempt:2012fy}).
\vspace{-2mm}
}
\end{figure}

\section{\label{4th}The Regge trajectories}

The masses of light-quark baryons fall onto Regge trajectories. Figure~\ref{pic:regge}
shows the Regge trajectory of $\Delta$ baryons; plotted is the squared baryon mass $M^2$ versus
the total angular momentum $J$. The four states $\Delta(1232)3/2^+$,  $\Delta(1950)7/2^+$,
$\Delta(2420)11/2^+$, and $\Delta(2950)15/2^+$ -- all having $J=L+S$ with $L=0, .. ,4$ and $S=3/2$ --
are compatible with a linear trajectory. This trajectory is compared with the mesonic trajectory, again
for mesons with $J=L+S$  but $S=1$ and for even and odd angular momenta. (Note that
the negative parity $\Delta(1700)3/2^-$, $\Delta(2200)7/2^-$ and likely $\Delta(2750)11/2^-$ have
spin $S=1/2$. Nevertheless, they fall onto the trajectory  shown in Fig.~\ref{pic:regge} when
the orbital angular momentum $L$ instead of $J$ is considered.)

For $\Sigma$ resonances, there are only two states that can be considered at the moment: $\Sigma(1385)3/2^+$
and $\Sigma(2030)7/2^+$. Their squared-mass difference suggests an identical slope as the one for $\Delta$ states.
Nevertheless, it would be important to increase our knowledge on high-mass $\Sigma$ resonances.

The $\Lambda$ Regge trajectory could be extracted from an analysis of $K_L n$ interactions. Here, $\Lambda$
resonances in SU(3) singlet and octet and $\Sigma$ resonances in SU(3) octet and decuplet contribute. Compared
to $K_L p$ interactions, this is certainly a more complicated task.
\section{\label{5th}Cascade baryons}
There is only one single $\Xi$ resonance, $\Xi(1820)3/2^-$, with known spin-parity. It is
290\,MeV heavier than the SU(3) decuplet ground state, $\Xi(1530)3/2^+$. The difference corresponds to
the mass gap between $N(1520)3/2^-$ and $\Delta(1232)$ $3/2^+$. Hence, very likely, $\Xi(1820)3/2^-$
is a SU(3) octet state and the first orbital excitation of the $\Xi$. Otherwise, the $\Xi$ resonances
are an uncharted territory.

\begin{figure}[ph]
\bc
\begin{tabular}{cc}
\includegraphics[width=0.23\textwidth,height=0.15\textwidth]{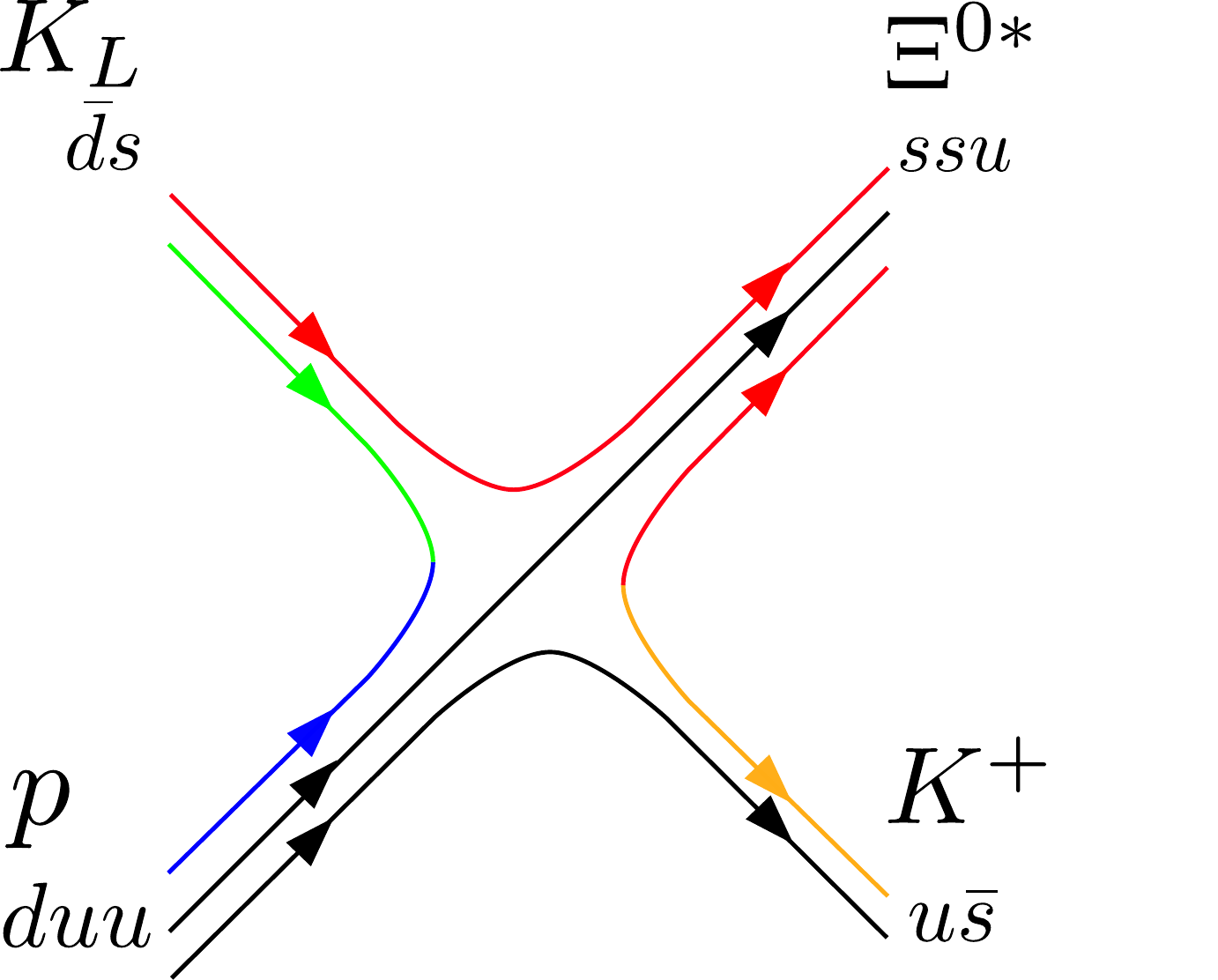}&\hspace{-4mm}
\includegraphics[width=0.23\textwidth,height=0.15\textwidth]{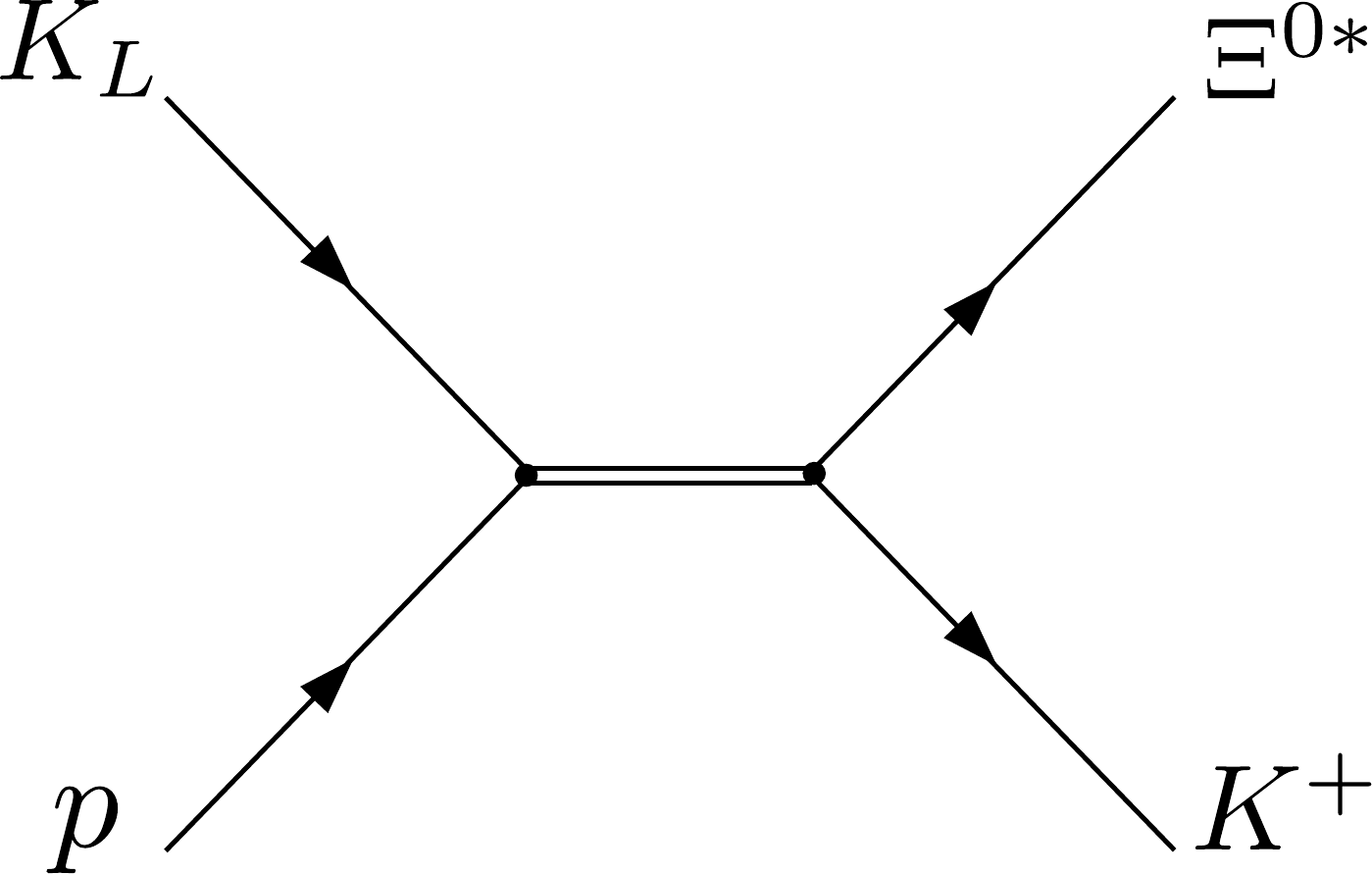}
\end{tabular}
\ec
\caption{\label{pic:Xi-prod}Left: Quark flow diagram for the  production of 
a $\Xi^{0*}$ resonance via $K_L p\to K^+ \Xi^{0*}$.
$s$-quarks in red, $\bar s$ in orange, $d$-quarks in blue, $\bar d$ in green, $u$-quarks in black.
Right: Hadron representation of the scattering process.\vspace{-2mm}
}
\end{figure}
Figure~\ref{pic:Xi-prod} shows the flux diagram for the production of $\Xi^{0*}$ resonances.
The intermediate state is a $\Sigma^+$ excitation. 
In the first searches for states, the decay of $\Xi$ resonances into $\pi\Xi$ with a subsequent
weak $\Xi$ decay should greatly reduce the background.
\section{\label{6th}Summary}

The planned $K_L$ beamline at JLab in connection with the GlueX experiment provides a powerful tool to study
$K N$ and $\bar KN$ interactions. In formation experiments, $\Sigma$ resonances can be studied
using a proton target; a deuteron target makes $\Lambda$ resonances accessible. In this paper, we emphasize
the highlights of production experiments with several particles in the final state. We propose to determine the disputed SU(3) structure of
$\Lambda(1405)$ and propose a method how to search for members of the missing 20-plet, in particular
for $\Lambda$ states in the $^41$ multiplet, in a cascade process. Further, we suggest to search for light-quark
pentaquarks with quantum numbers that are incompatible with a $qqq$ interpretation. 
The study can be extended to identify resonances of hyperons with two units of strangeness.\vspace{3mm}

{\it
This work was supported by the \textit{Deutsche Forschungsgemeinschaft} (SFB/TR110)%and the \textit{Russian Science Foundation} (RSF 16-12-10267)
.
}


\begin{thebibliography}{99}
\bibitem{Tanabashi:2018oca}
  M.~Tanabashi {\it et al.} [Particle Data Group],
  %``Review of Particle Physics,''
  Phys.\ Rev.\ D {\bf 98}, no. 3, 030001 (2018).

\bibitem{Zhang:2013cua}
  H.~Zhang, J.~Tulpan, M.~Shrestha and D.M.~Manley,
  %``Partial-wave analysis of $N\bar K$ scattering reactions,''
  Phys.\ Rev.\ C {\bf 88}, no. 3, 035204 (2013).

\bibitem{Zhang:2013sva}
  H.~Zhang, J.~Tulpan, M.~Shrestha and D.M.~Manley,
  %``Multichannel parametrization of $N\bar K$ scattering amplitudes and extraction of resonance parameters,''
  Phys.\ Rev.\ C {\bf 88}, no. 3, 035205 (2013).

\bibitem{Fernandez-Ramirez:2015tfa}
  C.~Fernandez-Ramirez, I.V.~Danilkin, D.M.~Manley, V.~Mathieu and A.P.~Szczepaniak,
  %``Coupled-channel model for $\bar{K}N$ scattering in the resonant region,''
  Phys.\ Rev.\ D {\bf 93}, no. 3, 034029 (2016).

\bibitem{Kamano:2014zba}
  H.~Kamano, S.X.~Nakamura, T.-S.H.~Lee and T.~Sato,
  %``Dynamical coupled-channels model of K$^−$p reactions: Determination of partial-wave amplitudes,''
  Phys.\ Rev.\ C {\bf 90}, no. 6, 065204 (2014).

\bibitem{Kamano:2015hxa}
  H.~Kamano, S.X.~Nakamura, T.-S.H.~Lee and T.~Sato,
  %``Dynamical coupled-channels model of $K^- p$ reactions. II. Extraction of $\Lambda^*$ and $\Sigma^*$ hyperon resonances,''
  Phys.\ Rev.\ C {\bf 92}, no. 2, 025205 (2015)
  Erratum: [Phys.\ Rev.\ C {\bf 95}, no. 4, 049903 (2017)].

\bibitem{Matveev:2019igl}
M.~Matveev, A.~Sarantsev, V.~Nikonov, A.~Anisovich, U.~Thoma and E.~Klempt,
%``Hyperon I: Partial-wave amplitudes for K$^{-}$p scattering,''
Eur. Phys. J. A \textbf{55} (2019) no.10, 179.
  %%CITATION = ARXIV:1907.03645;%%

\bibitem{Sarantsev:2019xxm}
A.~Sarantsev, M.~Matveev, V.~Nikonov, A.~Anisovich, U.~Thoma and E.~Klempt,
%``Hyperon II: Properties of excited hyperons,''
Eur. Phys. J. A \textbf{55} (2019) no.10, 180.

\bibitem{Isgur:1978wd}
  N.~Isgur and G.~Karl,
  %``Positive Parity Excited Baryons in a Quark Model with Hyperfine Interactions,''
  Phys.\ Rev.\ D {\bf 19}, 2653 (1979).
  Erratum: [Phys.\ Rev.\ D {\bf 23}, 817 (1981)].

 \bibitem{Loring:2001ky}
  U.~L\"oring, B.~C.~Metsch and H.~R.~Petry,
  %``The Light baryon spectrum in a relativistic quark model with instanton induced quark forces: The Strange baryon spectrum,''
  Eur.\ Phys.\ J.\ A {\bf 10}, 447 (2001).

\bibitem{Kuznetsov:2017ayk}
  V.~Kuznetsov {\it et al.},
  %``New narrow $N(1685)$ and $N(1726)$? Remarks on the interpretation of the neutron anomaly as an interference phenomenon,''
  Pisma Zh.\ Eksp.\ Teor.\ Fiz.\  {\bf 105}, no. 10, 591 (2017).

\bibitem{Aaij:2019vzc}
  R.~Aaij {\it et al.} [LHCb Collaboration],
  %``Observation of a narrow pentaquark state, $P_c(4312)^+$, and of two-peak structure of the $P_c(4450)^+$,''
  Phys.\ Rev.\ Lett.\  {\bf 122}, no. 22, 222001 (2019).

\bibitem{Ali:2019lzf}
A.~Ali \textit{et al.} [GlueX],
%``First Measurement of Near-Threshold J/ψ Exclusive Photoproduction off the Proton,''
Phys. Rev. Lett. \textbf{123}, no.7, 072001 (2019). 

\bibitem{Amaryan:2017ldw}
  S.~Adhikari {\it et al.} [GlueX Collaboration],
  ``Strange Hadron Spectroscopy with a Secondary $K_L$ Beam at GlueX,''
Proposal to the JLab Program Advisory Committee, PAC47 (2019).
  arXiv:1707.05284 [hep-ex].

\bibitem{Ketzer:2019wmd}
B.~Ketzer, B.~Grube and D.~Ryabchikov,
``Light-Meson Spectroscopy with COMPASS,''
[arXiv:1909.06366 [hep-ex]].

\bibitem{Sako:2013prop}
K.H. Hicks and H. Sako,
``P45: 3-Body Hadronic Reactions for New Aspects of Baryon Spectroscopy'',
Proposal for J-PARC E45 (2013).

\bibitem{Iazzi:2016fzb}
  F.~Iazzi [PANDA Collaboration],
  %``The PANDA physics program: Strangeness and more,''
  AIP Conf.\ Proc.\  {\bf 1743}, 050006 (2016).

\bibitem{Guzey:2005rx}
  V.~Guzey and M.V.~Polyakov,
  %``SU(3) systematization of baryons: Theoretical methods and mixing with the antidecuplet,''
  Annalen Phys.\  {\bf 13}, 673 (2004).

\bibitem{Guzey:2005vz}
  V.~Guzey and M.V.~Polyakov,
  %``SU(3) systematization of baryons,''
  hep-ph/0512355.

\bibitem{Klempt:2012fy}
E.~Klempt and B.~Metsch,
%``Multiplet classification of light-quark baryons,''
Eur. Phys. J. A \textbf{48} (2012), 127.

\bibitem{Alston:1961zzd}
  M.~H.~Alston {\it et al.}, %L.~W.~Alvarez, P.~Eberhard, M.~L.~Good, W.~Graziano, H.~K.~Ticho and S.~G.~Wojcicki,
  %``Study of Resonances of the Sigma-pi System,''
  Phys.\ Rev.\ Lett.\  {\bf 6}, 698 (1961).

\bibitem{Isgur:1978xj}
  N.~Isgur and G.~Karl,
  %``P Wave Baryons in the Quark Model,''
  Phys.\ Rev.\ D {\bf 18}, 4187 (1978).

\bibitem{Tripp:1969rd}
  R.D.~Tripp, R.O.~Bangerter, A.~Barbaro-Galtieri and T.S.~Mast,
  %``Direct evidence for the multiplet assignments of lambda(1520) and lambda(1405),''
  Phys.\ Rev.\ Lett.\  {\bf 21}, 1721 (1968).

\bibitem{Moriya:2014kpv}
  K.~Moriya {\it et al.} [CLAS Collaboration],
  %``Spin and parity measurement of the Lambda(1405) baryon,''
  Phys.\ Rev.\ Lett.\  {\bf 112}, 082004 (2014).

\bibitem{Kaiser:1996js}
  N.~Kaiser, T.~Waas and W.~Weise,
  %``SU(3) chiral dynamics with coupled channels: Eta and kaon photoproduction,''
  Nucl.\ Phys.\ A {\bf 612}, 297 (1997).

\bibitem{Oller:2000fj}
  J.~A.~Oller and U.-G.~Mei\ss ner,
  %``Chiral dynamics in the presence of bound states: Kaon nucleon interactions revisited,''
  Phys.\ Lett.\ B {\bf 500}, 263 (2001).

\bibitem{Jido:2003cb}
  D.~Jido, J.~A.~Oller, E.~Oset, A.~Ramos and U.-G.~Mei\ss ner,
  %``Chiral dynamics of the two Lambda(1405) states,''
  Nucl.\ Phys.\ A {\bf 725}, 181 (2003).

\bibitem{Oset:2001cn}
  E.~Oset, A.~Ramos and C.~Bennhold,
  %``Low lying S = -1 excited baryons and chiral symmetry,''
  Phys.\ Lett.\ B {\bf 527}, 99 (2002)
   Erratum: [Phys.\ Lett.\ B {\bf 530}, 260 (2002)]

\bibitem{Cieply:2009ea}
  A.~Cieply and J.~Smejkal,
  %``Separable potential model for K- N interactions at low energies,''
  Eur.\ Phys.\ J.\ A {\bf 43}, 191 (2010).

\bibitem{Cieply:2011nq}
  A.~Cieply and J.~Smejkal,
  %``Chirally motivated $\bar{K}N$ amplitudes for in-medium applications,''
  Nucl.\ Phys.\ A {\bf 881}, 115 (2012).

\bibitem{Ikeda:2011pi}
  Y.~Ikeda, T.~Hyodo and W.~Weise,
  %``Improved constraints on chiral SU(3) dynamics from kaonic hydrogen,''
  Phys.\ Lett.\ B {\bf 706}, 63 (2011).

\bibitem{Ikeda:2012au}
  Y.~Ikeda, T.~Hyodo and W.~Weise,
  %``Chiral SU(3) theory of antikaon-nucleon interactions with improved threshold constraints,''
  Nucl.\ Phys.\ A {\bf 881}, 98 (2012).

 \bibitem{Guo:2012vv}
  Z.~H.~Guo and J.~A.~Oller,
  %``Meson-baryon reactions with strangeness -1 within a chiral framework,''
  Phys.\ Rev.\ C {\bf 87}, no. 3, 035202 (2013).

\bibitem{Mai:2012dt}
  M.~Mai and U.-G.~Mei\ss ner,
  %``New insights into antikaon-nucleon scattering and the structure of the Lambda(1405),''
  Nucl.\ Phys.\ A {\bf 900}, 51  (2013).
%

\bibitem{Mai:2014xna}
  M.~Mai and U.-G.~Mei\ss ner,
  %``Constraints on the chiral unitary $\bar KN$ amplitude from $\pi\Sigma K^+$ photoproduction data,''
  Eur.\ Phys.\ J.\ A {\bf 51},  30 (2015).

\bibitem{Roca:2013av}
  L.~Roca and E.~Oset,
  %``?(1405) poles obtained from $?^0?^0$ photoproduction data,''
  Phys.\ Rev.\ C {\bf 87},  055201 (2013).

\bibitem{Roca:2013cca}
  L.~Roca and E.~Oset,
  %``Isospin 0 and 1 resonances from $\pi \Sigma$ photoproduction data,''
  Phys.\ Rev.\ C {\bf 88}, 055206 (2013).

\bibitem{Miyahara:2018onh}
  K.~Miyahara, T.~Hyodo and W.~Weise,
  %``Construction of a local $\bar K N-\pi \Sigma-\pi \Lambda$ potential and composition of the $\Lambda(1405)$,''
  Phys.\ Rev.\ C {\bf 98}, 025201 (2018).

\bibitem{Feijoo:2018den}
  A.~Feijoo, V.~Magas and A.~Ramos,
  %``$S=-1$ meson-baryon interaction and the role of isospin filtering processes,''
  Phys.\ Rev.\ C {\bf 99}, no. 3, 035211 (2019).

\bibitem{Cieply:2016jby}
  A.~Cieply, M.~Mai, U.-G.~Mei\ss ner and J.~Smejkal,
  %``On the pole content of coupled channels chiral approaches used for the $\bar{K}N$ system,''
  Nucl.\ Phys.\ A {\bf 954}, 17 (2016).

\bibitem{LeviSetti69}R. Levi-Setti,
in: {\it Proceedings of the Lund International Conference on Elementary Particles,
Lund, 1969, p.~339}, G. von Dardel (ed.); Berlingska Boktryckeriet, Lund, Sweden 1969.

\bibitem{Moriya:2013hwg}
  K.~Moriya {\it et al.} [CLAS Collaboration],
  %``Differential Photoproduction Cross Sections of the $\Sigma^0(1385)$, $\Lambda(1405)$, and $\Lambda(1520)$,''
  Phys.\ Rev.\ C {\bf 88}, 045201 (2013).
  Addendum: [Phys.\ Rev.\ C {\bf 88}, 049902 (2013)].

\bibitem{Humphrey:1962zz}
  W.~E.~Humphrey and R.~R.~Ross,
  %``Low-Energy Interactions of K- Mesons in Hydrogen,''
  Phys.\ Rev.\  {\bf 127}, 1305 (1962).

\bibitem{Watson:1963zz}
  M.~B.~Watson, M.~Ferro-Luzzi and R.~D.~Tripp,
  %``Analysis of Y-0* (1520) and Determination of the Sigma Parity,''
  Phys.\ Rev.\  {\bf 131}, 2248 (1963).

\bibitem{Sakitt:1965kh}
  M.~Sakitt, T.~B.~Day, R.~G.~Glasser, N.~Seeman, J.~H.~Friedman, W.~E.~Humphrey and R.~R.~Ross,
  %``Low-energy K- Meson Interactions In Hydrogen,''
  Phys.\ Rev.\  {\bf 139}, B719 (1965).

\bibitem{Ciborowski:1982et}
  J.~Ciborowski {\it et al.},
  %``Kaon Scattering And Charged Sigma Hyperon Production In K- P Interactions Below 300-mev/c,''
  J.\ Phys.\ G {\bf 8}, 13 (1982).

\bibitem{Mast:1975pv}
T.~S.~Mast, M.~Alston-Garnjost, R.~O.~Bangerter, A.~S.~Barbaro-Galtieri, F.~T.~Solmitz and R.~D.~Tripp,
%``Elastic, Charge Exchange, and Total K- p Cross-Sections in the Momentum Range 220-MeV/c to 470-MeV/c,''
Phys.\ Rev.\ D {\bf 14}, 13 (1976).

\bibitem{Prakhov:2004ri}
  S.~Prakhov {\it et al.},
  %``Reaction K- p ---> pi0 pi0 lambda from pK- = 514-MeV/c to 750-MeV/c,''
  Phys.\ Rev.\ C {\bf 69}, 042202 (2004).

\bibitem{Prakhov:2004an}
  S.~Prakhov {\it et al.},
  %``K- p ---> pi0 pi0 sigma0 at p(K)- = 514-MeV/c to 750-MeV/c mev and comparison with other pi0 pi0 production,''
  Phys.\ Rev.\ C {\bf 70}, 034605 (2004).

\bibitem{Hemingway:1984pz}
  R.~J.~Hemingway,
  %``Production of $\Lambda(1405)$ in $K^- p$ Reactions at 4.2-{GeV}/$c$,''
  Nucl.\ Phys.\ B {\bf 253}, 742 (1985).

\bibitem{Tovee:1971ga}
  D.~N.~Tovee {\it et al.},
  %``Some properties of the charged sigma hyperons,''
  Nucl.\ Phys.\ B {\bf 33}, 493 (1971).

\bibitem{Nowak:1978au}
  R.~J.~Nowak {\it et al.},
  %``Charged Sigma Hyperon Production by K- Meson Interactions at Rest,''
  Nucl.\ Phys.\ B {\bf 139}, 61 (1978).

\bibitem{Bazzi:2011zj}
  M.~Bazzi {\it et al.} [SIDDHARTA Collaboration],
  %``A New Measurement of Kaonic Hydrogen X-rays,''
  Phys.\ Lett.\ B {\bf 704}, 113 (2011).

\bibitem{Bazzi:2012eq}
  M.~Bazzi {\it et al.},
  %``Kaonic hydrogen X-ray measurement in SIDDHARTA,''
  Nucl.\ Phys.\ A {\bf 881}, 88 (2012).

\bibitem{Armenteros:1970eg}
  R.~Armenteros {\it et al.},
%  ``$K^-$ p cross-sections from 440 to 800 Mev/c,''
  Nucl.\ Phys.\ B {\bf 21}, 15 (1970).

\bibitem{Anisovich:2020tbd}
A.V. Anisovich, A.V. Sarantsev, V.A. Nikonov, V. Burkert,
R. Schumacher, E. Klempt, ``Hyperon III:  $K^-p - \pi\Sigma$ coupled-channel dynamics in the $\Lambda(1405)$ mass region''
accepted for publication in EPJA.

\bibitem{Sokhoyan:2015fra}
  V.~Sokhoyan {\it et al.} [CBELSA/TAPS Collaboration],
  %``High-statistics study of the reaction $\gamma p\to p\;2\pi^0$,''
  Eur.\ Phys.\ J.\ A {\bf 51}, no. 8, 95 (2015).

\bibitem{Thiel:2015kuc}
  A.~Thiel {\it et al.} [CBELSA/TAPS Collaboration],
  %``Three-body nature of $N^{\bf *}$ and $\Delta^*$ resonances from sequential decay chains,''
  Phys.\ Rev.\ Lett.\  {\bf 114}, no. 9, 091803 (2015).

\bibitem{Skyrme:1961vq}
  T.~H.~R.~Skyrme,
  %``A Nonlinear field theory,''
  Proc.\ Roy.\ Soc.\ Lond.\ A {\bf 260}, 127 (1961).

\bibitem{Witten:1983tw}
  E.~Witten,
  %``Global Aspects of Current Algebra,''
  Nucl.\ Phys.\ B {\bf 223}, 422 (1983).

\bibitem{Chemtob:1985ar}
  M.~Chemtob,
  %``Skyrme Model of Baryon Octet and Decuplet,''
  Nucl.\ Phys.\ B {\bf 256}, 600 (1985).

\bibitem{Walliser:1992vx}
  H.~Walliser,
  %``The SU(n) Skyrme model,''
  Nucl.\ Phys.\ A {\bf 548}, 649 (1992).

\bibitem{Diakonov:1997mm}
  D.~Diakonov, V.~Petrov and M.~V.~Polyakov,
  %``Exotic anti-decuplet of baryons: Prediction from chiral solitons,''
  Z.\ Phys.\ A {\bf 359}, 305 (1997).

\bibitem{Diakonov:2003jj}
  D.~Diakonov and V.~Petrov,
  %``Where are the missing members of the baryon anti-decuplet?,''
  Phys.\ Rev.\ D {\bf 69}, 094011 (2004).

\bibitem{Nakano:2003qx}
  T.~Nakano {\it et al.} [LEPS Collaboration],
  %``Evidence for a narrow S = +1 baryon resonance in photoproduction from the neutron,''
  Phys.\ Rev.\ Lett.\  {\bf 91}, 012002 (2003).

\bibitem{Barmin:2003vv}
  V.~V.~Barmin {\it et al.} [DIANA Collaboration],
  %``Observation of a baryon resonance with positive strangeness in K+ collisions with Xe nuclei,''
  Phys.\ Atom.\ Nucl.\  {\bf 66}, 1715 (2003)
  [Yad.\ Fiz.\  {\bf 66}, 1763 (2003)].

\bibitem{Barth:2003es}
  J.~Barth {\it et al.} [SAPHIR Collaboration],
  %``Evidence for the positive strangeness pentaquark Theta+ in photoproduction with the SAPHIR detector at ELSA,''
  Phys.\ Lett.\ B {\bf 572}, 127 (2003).

\bibitem{Stepanyan:2003qr}
  S.~Stepanyan {\it et al.} [CLAS Collaboration],
  %``Observation of an exotic S = +1 baryon in exclusive photoproduction from the deuteron,''
  Phys.\ Rev.\ Lett.\  {\bf 91}, 252001 (2003).

\bibitem{Dzierba:2004db}
  A.~R.~Dzierba, C.~A.~Meyer and A.~P.~Szczepaniak,
  %``Reviewing the evidence for pentaquarks,''
  J.\ Phys.\ Conf.\ Ser.\  {\bf 9}, 192 (2005).

\bibitem{Danilov:2007bp}
  M.~Danilov and R.~Mizuk,
  %``Experimental review on pentaquarks,''
  Phys.\ Atom.\ Nucl.\  {\bf 71}, 605 (2008).

\bibitem{Liu:2014yva}
  T.~Liu, Y.~Mao and B.~Q.~Ma,
  %``Present status on experimental search for pentaquarks,''
  Int.\ J.\ Mod.\ Phys.\ A {\bf 29}, no. 13, 1430020 (2014).

\bibitem{Kuznetsov:2014aka}
  V.~Kuznetsov,
  ``New narrow N*(1685) resonance: review of observations,''
  EPJ Web Conf.\  {\bf 73}, 04020 (2014).

\bibitem{Kuznetsov:2015nla}
  V.~Kuznetsov {\it et al.},
  %``Evidence for narrow resonant structures at W≈1.68GeV and W≈1.72GeV in real Compton scattering off the proton,''
  Phys.\ Rev.\ C {\bf 91}, no. 4, 042201 (2015).

\bibitem{Gridnev:2016dba}
  A.~Gridnev {\it et al.} [EPECUR Collaboration],
  %``Search for narrow resonances in $\pi p$ elastic scattering from the EPECUR experiment,''
  Phys.\ Rev.\ C {\bf 93}, no. 6, 062201 (2016).

\bibitem{Alt:2003vb}
  C.~Alt {\it et al.} [NA49 Collaboration],
  %``Observation of an exotic S = -2, Q = -2 baryon resonance in proton proton collisions at the CERN SPS,''
  Phys.\ Rev.\ Lett.\  {\bf 92}, 042003 (2004).

\bibitem{Anisovich:2015tla}
A.~Anisovich, E.~Klempt, B.~Krusche, V.~Nikonov, A.~Sarantsev, U.~Thoma and D.~Werthmüller,
%``Interference phenomena in the $J^P=1/2^-$ wave in $\eta$ photoproduction,''
Eur. Phys. J. A \textbf{51} (2015) no.6, 72.
%34 citations counted in INSPIRE as of 12 May 2020

\bibitem{Werthmuller:2015owc}
D.~Werthmüller, L.~Witthauer, D.~Glazier and B.~Krusche,
%``Comment on "Evidence for Narrow Resonant Structures at $W\approx$ 1.68 GeV and $W\approx$ 1.72 GeV in Real Compton Scattering Off the Proton",''
Phys. Rev. C \textbf{92} (2015) no.6, 069801.
%11 citations counted in INSPIRE as of 12 May 2020

\bibitem{Witthauer:2017get}
L.~Witthauer \textit{et al.} [A2],
%``Insight into the Narrow Structure in η Photoproduction on the Neutron from Helicity-Dependent Cross Sections,''
Phys. Rev. Lett. \textbf{117} (2016) no.13, 132502.
%25 citations counted in INSPIRE as of 12 May 2020

\bibitem{Anisovich:2017xqg}
  A.~V.~Anisovich, V.~Burkert, E.~Klempt, V.~A.~Nikonov, A.~V.~Sarantsev and U.~Thoma,
  %``Scrutinizing the evidence for N(1685),''
  Phys.\ Rev.\ C {\bf 95}, no. 3, 035211 (2017).

\bibitem{Kuznetsov:2017xgu}
V.~Kuznetsov, F.~Mammoliti, F.~Tortorici, V.~Bellini, V.~Brio, A.~Gridnev, N.~Kozlenko, G.~Russo, M.~Sperduto, V.~Sumachev and C.~Sutera,
%``Observation of Narrow $N^+(1685)$ and $N^0(1685)$ Resonances in $\gamma N \to \eta \pi N$ Reactions,''
JETP Lett. \textbf{106} (2017) no.11, 693-699.
%9 citations counted in INSPIRE as of 12 May 2020

\bibitem{Kuznetsov:2017qmo}
V.~Kuznetsov, V.~Bellini, V.~Brio, A.~Gridnev, N.~Kozlenko, F.~Mammoliti, F.~Tortorici, M.~Polyakov, G.~Russo, M.~Sperduto, V.~Sumachev and C.~Sutera,
%``New Narrow $N(1685)$ and $N(1726)$? Remarks on the Interpretation of the Neutron Anomaly as an Interference Phenomenon,''
JETP Lett. \textbf{105} (2017) no.10, 625-630.
%7 citations counted in INSPIRE as of 12 May 2020

\bibitem{Kuznetsov:2018dcd}
  V.~A.~Kuznetsov and M.~V.~Polyakov,
 ``Large violation of the flavour SU(3) symmetry in $\eta$MAID2018 isobar model,''
 (unpublished) arXiv:1810.07713 [nucl-th].

\bibitem{Goeke:2009ae}
  K.~Goeke, M.~V.~Polyakov and M.~Praszalowicz,
  %``On strange SU(3) partners of Theta+,''
  Acta Phys.\ Polon.\ B {\bf 42}, 61 (2011).
\end{thebibliography}
\end{document}